\documentclass[
aip,
 amsmath,amssymb,
reprint,%
]{revtex4-1}

\usepackage{graphicx}
\usepackage{dcolumn}
\usepackage{bm}
\usepackage{float} 
\usepackage{makecell}
\usepackage[table]{xcolor}
\usepackage[utf8]{inputenc}
\usepackage[T1]{fontenc}
\usepackage{mathptmx}
\usepackage{etoolbox}
\usepackage{multirow}
\usepackage{xcolor}
\makeatletter
\def\@email#1#2{%
 \endgroup
 \patchcmd{\titleblock@produce}
  {\frontmatter@RRAPformat}
  {\frontmatter@RRAPformat{\produce@RRAP{*#1\href{mailto:#2}{#2}}}\frontmatter@RRAPformat}
  {}{}
}%
\makeatother
\begin{document}
\preprint{AIP/123-QED}
\title[Hybrid Parametric PINNs for Incompressible Flows] {Sparse-Supervised Hybrid Parameterized Physics-Informed Neural Networks for Incompressible Flows Across Reynolds Numbers}

\author{A. Jangir}
\thanks{Electronic mail: adityajangir873@gmail.com}
\affiliation{Department of Energy Science and Engineering, IIT Delhi}

\author{R. Clements}
\affiliation{Department of Engineering, FESE, University of Exeter}

\author{R. Goyal}
\affiliation{Department of Energy Science and Engineering, IIT Delhi}

\author{G. Tabor}
\thanks{Electronic mail: g.r.tabor@exeter.ac.uk}
\affiliation{Department of Engineering, FESE, University of Exeter}

\date{\today}

\begin{abstract}
Physics-informed neural networks (PINNs) provide a mesh-free framework for solving partial differential equations by embedding governing physical laws directly into neural-network training. Recent studies have shown that parameterized PINNs can learn families of Navier-Stokes solutions across Reynolds numbers by treating governing parameters as additional network inputs. However, the accuracy of data-free PINNs deteriorates significantly in convection-dominated high-Reynolds-number flows because of optimization stiffness and sharp multiscale flow structures.

Motivated by these limitations, the present study investigates a sparse-supervised hybrid parameterized PINNs framework for incompressible Navier-Stokes flows, emphasizing regime-aware learning, localized Reynolds-number supervision, and data-efficient correction of high-Reynolds-number failure modes. The Reynolds number is incorporated explicitly as a network input, enabling a single model to learn a continuous parametric flow manifold across different operating conditions.

The methodology is first demonstrated for two-dimensional lid-driven cavity flow and further assessed using backward-facing step flow to evaluate generalizability to a qualitatively different separated-flow configuration. For low Reynolds numbers, data-free PINNs accurately recover velocity and pressure fields using only governing equations and boundary conditions. At higher Reynolds numbers, where convection-dominated effects degrade physics-only PINNs, a hybrid framework combining transfer learning and localized sparse computational fluid dynamics (CFD) supervision is introduced.

Although the training range spans \(500 < Re < 1000\), supervised CFD data are provided only within the restricted interval \(750 < Re < 850\). Supervision is additionally limited to only \(3\%-20\%\) of the computational points to quantify minimum data requirements. The results demonstrate that approximately \(5\%\) supervised data are sufficient to recover accurate solutions with high predictive fidelity.

Comparisons with CFD simulations demonstrate strong agreement in velocity, pressure, vorticity, and reattachment characteristics across interpolation and limited extrapolation regimes. The study highlights a practical physics-first hybridization strategy in which sparse data are introduced only when data-free PINNs become insufficient, providing a data-efficient framework for incompressible flows across Reynolds numbers.
\end{abstract}

\maketitle

\section{\label{sec:level1}INTRODUCTION:\protect\\  }

The lid-driven cavity (LDC) flow is one of the most fundamental benchmark problems in fluid dynamics and has long served as a reference configuration for studying incompressible viscous flows and validating numerical methods. Owing to its geometric simplicity, well-defined boundary conditions, and strong sensitivity to the Reynolds number, the LDC flow remains a canonical test case for assessing numerical accuracy, stability, and convergence properties of Navier-Stokes solvers~\cite{Shankar2000}. Despite its simplicity, the flow exhibits rich physical behavior, including primary recirculation, secondary corner vortices, and pronounced changes in flow topology as the Reynolds number increases, making it particularly suitable for investigating parametric flow dependence.

Seminal benchmark solutions for the two-dimensional incompressible LDC flow were provided by Ghia \textit{et al.}~\cite{Ghia1982}, who employed a multigrid finite-difference method to compute highly resolved velocity and streamfunction fields over a wide range of Reynolds numbers. Their centerline velocity profiles and vortex strength data have since become standard references for validating CFD solvers. Botella and Peyret~\cite{Botella1998} later extended this work using a high-order Chebyshev collocation method with a regularized lid velocity profile, producing spectral benchmark solutions that remain definitive references for numerical accuracy assessment. Further advances were reported by Erturk \textit{et al.}~\cite{Erturk2005,Erturk2006}, who introduced compact high-order finite-difference schemes and provided detailed analyses of convergence behavior and secondary vortical structures at higher Reynolds numbers. These studies firmly established the LDC flow as a stringent benchmark for incompressible Navier-Stokes solvers.

At low Reynolds numbers, the flow is steady and laminar, characterized by a dominant central vortex and weaker secondary vortices near the corners. As the Reynolds number increases, nonlinear convective effects become increasingly important, leading to stronger velocity gradients, enhanced secondary structures, and, in three-dimensional configurations, transition toward unsteady and turbulent flow. These features have made the LDC problem a cornerstone in CFD, numerical analysis, and, more recently, data-driven and machine-learning-based approaches for fluid flows, including PINNs and hybrid CFD-ML methods~\cite{Duraisamy2018arxiv,Brunton2020,Cai2024PoF}.

In contrast to conventional discretization-based numerical methods, Raissi, Perdikaris, and Karniadakis~\cite{Raissi2019} introduced the physics-informed neural networks framework, in which the governing equations and boundary conditions are embedded directly into the neural-network training process through physics-based loss functions. This mesh-free framework enables the solution of partial differential equations while enforcing physical consistency and has attracted significant attention for applications involving fluid mechanics, inverse problems, turbulence modeling, and multiphysics systems~\cite{Raissi2018HiddenFluidMechanics,Yang2025PoF_ROM,Roy2025PoF,Prieto2025PoF}.

Building upon the original PINNs formulation, numerous extensions have subsequently been proposed to improve convergence, robustness, and predictive accuracy for Navier--Stokes problems. Kharazmi, Zhang, and Karniadakis~\cite{Kharazmi2021} introduced the hp-variational physics-informed neural networks (hp-VPINNs) framework, which combines domain decomposition and weak formulations to alleviate optimization stiffness at elevated Reynolds numbers. Li and Feng~\cite{Li2022} proposed adaptive loss-weighting strategies for balancing competing residual terms during training, while Chen and Zhao~\cite{Chen2025PoF107129} improved the performance of hard-constrained gradient-enhanced PINNs (gPINNs) using residual-based adaptive sampling strategies.

Recently, transfer learning and parameterized learning strategies have emerged as promising approaches for improving the efficiency and generalization capability of PINNs when solving families of related flow problems. Wang \textit{et al.}~\cite{Wang2025arxiv} demonstrated that transfer learning can significantly reduce training cost while maintaining accuracy across variations in boundary conditions, material properties, and geometries. Related ideas have also been explored in broader data-driven modeling of fluid flows~\cite{Geneva2020}, highlighting the potential of reusing learned representations across related flow configurations.

In parallel, increasing attention has been devoted to parameterized physics-informed neural networks (p-PINNs), in which physical or geometric parameters are explicitly incorporated into the network inputs to enable learning of generalized solution manifolds. Sun \textit{et al.}~\cite{Sun2020} introduced one of the earliest parameterized PINN formulations for PDEs involving varying material properties and governing parameters. More recently, parameterized PINNs have been increasingly applied to incompressible Navier-Stokes flows and fluid-dynamics surrogate modeling. Naderibeni \textit{et al.}~\cite{Naderibeni2024} demonstrated that a single PINN model can learn parametric solution manifolds of the Navier-Stokes equations by explicitly incorporating the Reynolds number into the network inputs. Yang \textit{et al.}~\cite{Yang2024} further explored data-driven physics-informed neural networks from a digital-twin perspective, emphasizing parametric learning and hybrid supervision strategies for fluid-flow prediction. Cao and Zhang~\cite{Cao2025} analyzed ill-conditioning mechanisms in PINNs and highlighted the increasing optimization difficulties associated with convection-dominated flows and high-Reynolds-number regimes. Additional recent studies have also investigated hybrid and fine-tuning strategies for cavity flows, including sparse-data-assisted PINNs and transfer-learning-based formulations for improving convergence and prediction robustness at elevated Reynolds numbers~\cite{SciRep2025,CMAME2025FineTune,MLDPINN2025}.

Despite these recent advances, several important challenges remain insufficiently explored for parameterized PINNs applied to incompressible flows. Existing studies primarily focus on demonstrating parametric interpolation capability using supervision distributed broadly throughout the parameter space or rely on comparatively dense supervised datasets. Moreover, although optimization difficulties of PINNs at elevated Reynolds numbers have been widely reported, the resulting deterioration is often discussed primarily from an optimization perspective, including spectral bias, loss imbalance, and Jacobian conditioning. Comparatively less attention has been devoted to understanding PINN behavior from a regime-dependent fluid-physics perspective, particularly regarding the transition from diffusion-dominated to convection-dominated flow regimes.

In practical CFD applications, high-fidelity labeled data are often available only for limited operating conditions due to the substantial computational expense associated with generating fully resolved simulations. Consequently, the capability of parameterized PINNs to recover accurate flow solutions outside sparsely supervised parameter regions remains poorly understood, particularly for convection-dominated incompressible flows at moderate and high Reynolds numbers.

In fluid dynamics, explicit treatment of governing non-dimensional parameters is especially important because these parameters dictate the balance of physical mechanisms in the flow. The Reynolds number controls the relative importance of convective and viscous transport in the Navier-Stokes equations and directly influences flow topology, vortex formation, shear-layer development, and optimization stiffness in PINNs training. Training separate PINNs for individual Reynolds numbers not only increases computational cost but also prevents the network from learning the continuous dependence of the solution manifold on governing physical parameters.

Motivated by these considerations, the present work systematically investigates a sparse-supervised hybrid parameterized PINNs framework for incompressible Navier-Stokes flows across Reynolds numbers. The Reynolds number is incorporated explicitly as a network input, enabling a unified neural representation of the parametric flow manifold. Unlike conventional hybrid PINN approaches that rely on supervision distributed throughout the parameter domain, the present framework adopts a physics-first strategy in which data-free PINNs are first employed to recover low-Reynolds-number solutions using only governing equations and boundary conditions. Sparse CFD supervision is subsequently introduced only when convection-dominated effects lead to deterioration of data-free PINNs performance at elevated Reynolds numbers.

The principal objective of the present study is therefore not to introduce parameterized PINNs themselves, which have already been explored in previous studies, but rather to systematically investigate regime-dependent behavior of parameterized PINNs and develop a practical hybridization strategy for convection-dominated incompressible flows. In particular, the study investigates how sparse localized CFD supervision and transfer learning can be used as corrective mechanisms when pure physics-based learning becomes insufficient at high Reynolds numbers. Although the training range spans a broader Reynolds-number interval, supervised CFD data are intentionally restricted to a localized Reynolds-number subdomain, thereby enabling assessment of the capability of physics-constrained learning to reconstruct unsupervised regions of the parametric solution space using minimal supervisory information.

The study additionally investigates several factors governing the performance and robustness of parameterized PINNs for incompressible flows. These include the influence of sampling strategies, collocation-point density, sparse-data fraction, gradient norm evolution during optimization, extrapolation degradation outside the training regime, and the increasing optimization stiffness encountered at elevated Reynolds numbers. The resulting analyses provide additional insight into the limitations and practical behavior of hybrid parameterized PINNs for convection-dominated Navier-Stokes problems.

To assess generalizability beyond the classical lid-driven cavity benchmark, the proposed framework is further evaluated using the backward-facing step (BFS) flow, which introduces separated shear layers, recirculation, and downstream reattachment. In contrast to the closed recirculating cavity configuration, the BFS problem provides a qualitatively different flow topology for assessing the transferability and robustness of the proposed methodology across distinct incompressible flow regimes.

Table~\ref{tab:comparison} summarizes the differences between representative recent parameterized PINN studies and the present work. In comparison with prior studies, the present work places particular emphasis on regime-aware PINN behavior, localized parameter-space supervision, sparse-data-assisted hybrid training, and systematic analysis of optimization behavior and parametric generalization for convection-dominated incompressible flows.

To the best of the authors' knowledge, systematic investigation of regime-dependent PINN behavior together with localized Reynolds-number supervision and sparse-supervised hybrid parameterized PINNs for convection-dominated incompressible flows has received comparatively limited attention in the existing literature.

The remainder of this paper is organized as follows. Section~II presents the computational methodology, including the CFD framework, PINNs formulation, neural-network architecture, loss functions, and training strategy. Section~III discusses the numerical results for lid-driven cavity and backward-facing step flows, including analyses of sampling strategies, collocation density, gradient evolution, sparse supervision, transfer learning, and comparisons with CFD simulations across different Reynolds numbers. Finally, Section~IV summarizes the main conclusions and outlines potential future research directions.

\begin{table*}[htbp]
\centering
\scriptsize
\caption{Comparison between representative recent PINN studies related to parametric incompressible flows and the present work.}
\label{tab:comparison}

\setlength{\tabcolsep}{2.5pt}

\renewcommand{\arraystretch}{1.15}

\resizebox{\textwidth}{!}{
\begin{tabular}{c c c c c c}
\hline

\textbf{Study} &
\textbf{Flow Configuration} &
\textbf{Re Input} &
\textbf{Supervision} &
\textbf{Transfer Learning} &
\textbf{Primary Focus} \\

\hline

Sun et al.~\cite{Sun2020}
&
Internal cardiovascular flows
&
Yes
&
Physics-only
&
No
&
Foundational parameterized PINNs for
Navier-Stokes systems
\\

Naderibeni et al.~\cite{Naderibeni2024}
&
2D cylinder flow
&
Yes
&
CFD-assisted
&
No
&
Parametric Navier-Stokes prediction across Reynolds numbers
\\

Yang et al.~\cite{Yang2024}
&
Lid-driven cavity flow
&
Yes
&
Hybrid multi-fidelity DD-PINNs
&
Partial
&
Adaptive sampling and digital-twin-oriented PINNs
\\

Cao and Zhang~\cite{Cao2025}
&
Cavity and M6 wing flows
&
No
&
Physics-only
&
No
&
PINN ill-conditioning and optimization stiffness analysis
\\

Lee et al.~\cite{SciRep2025}
&
High-Re cavity flow
&
No
&
Physics-only with re-initialization
&
No
&
Improving convergence of stiff PINNs by repeated parameter re-initialization
\\

Tsai et al.~\cite{MLDPINN2025}
&
Convection-diffusion and cavity flows
&
No
&
Multilevel dataset training
&
Yes
&
Stabilization of high-Re PINN training
\\

Present work
&
Cavity and backward-facing step flows
&
Yes
&
Physics-only + sparse CFD supervision
&
Yes
&
Regime-aware hybrid parameterized
PINNs with physics-first learning,
minimal-data correction, localized
parameter-space supervision,
\\

\hline
\end{tabular}
}
\end{table*}
\section{METHODOLOGY}
 \subsection{Computational Fluid Dynamics (CFD)}
\subsubsection{Problem Statement and Boundary Conditions}

The canonical two-dimensional lid-driven cavity problem is selected as the benchmark test case. The computational domain is a square cavity of dimension $1 \times 1$. The top wall (lid) moves with a uniform velocity $U_{\mathrm{lid}}$ in the positive $x$-direction, while the remaining three walls satisfy the no-slip boundary condition. This configuration generates a primary recirculating vortex inside the cavity, with additional secondary vortices developing near the corners as the Reynolds number increases.

The governing equations for incompressible viscous flow are the continuity and Navier-Stokes equations. In their general non-dimensional unsteady form, they are expressed as

\begin{equation}
\nabla \cdot \mathbf{u} = 0
\label{eq:continuity_vector}
\end{equation}

\begin{equation}
\frac{\partial \mathbf{u}}{\partial t}
+
(\mathbf{u} \cdot \nabla)\mathbf{u}
=
-\nabla p
+
\frac{1}{Re}\nabla^2 \mathbf{u}
\label{eq:NS_unsteady}
\end{equation}

where $\mathbf{u}=(u,v)$ denotes the velocity vector, $p$ is the pressure, and $Re=\frac{U_{\mathrm{lid}}L}{\nu}$ is the Reynolds number defined using the characteristic cavity length $L$, lid velocity $U_{\mathrm{lid}}$, and kinematic viscosity $\nu$.

Since the present study focuses on the steady-state lid-driven cavity flow, the transient term $\partial \mathbf{u}/\partial t$ vanishes after the flow reaches equilibrium. Consequently, the governing equations reduce to the steady incompressible Navier-Stokes equations:

\begin{equation}
(\mathbf{u} \cdot \nabla)\mathbf{u}
=
-\nabla p
+
\frac{1}{Re}\nabla^2 \mathbf{u}
\label{eq:NS_vector}
\end{equation}

The incompressible Navier-Stokes equations are written in non-dimensional form, where the Reynolds number acts as the primary governing parameter controlling the relative importance of convective and diffusive transport mechanisms. This formulation highlights the influence of Reynolds number on the resulting flow structure and solution behavior, consistent with classical fluid mechanics theory~\cite{White2006,Schlichting2017}.

\subsubsection{Validation of CFD results from existing literature}

CFD simulations were performed using the Open-source Field Operation And Manipulation (\texttt{OpenFOAM~12}) solver. High-fidelity CFD solutions obtained using \texttt{OpenFOAM~12} are used as reference data for validation. OpenFOAM has been extensively validated for incompressible laminar cavity flows and is widely used in benchmark CFD studies~\cite{Weller1998,Jasak2007}. A uniform computational grid consisting of $128 \times 128$ cells was employed, consistent with the grid resolution adopted in previous benchmark studies by Ghia \textit{et al.}~\cite{Ghia1982}, Botella and Peyret~\cite{Botella1998}, and Erturk \textit{et al.}~\cite{Erturk2005,Erturk2006}. Although Erturk \textit{et al.} reported that this resolution is generally sufficient for accurately resolving the dominant flow structures up to Reynolds numbers of approximately $10^{3}$, an additional validation was performed for the present study at $\mathrm{Re}=2000$ to ensure adequate spatial resolution for the higher Reynolds number case considered in the extrapolation analysis.

\begin{figure}[h]
    \centering
    \includegraphics[width=0.45\textwidth]{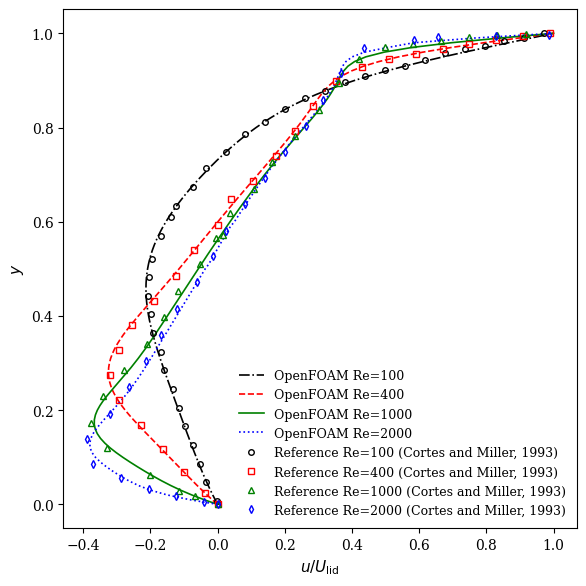}
    \caption{Comparison of horizontal velocity $u/U_{\mathrm{lid}}$ along the vertical centerline ($x=0.5$) between \texttt{OpenFOAM~12} results and reference data of Cortes and Miller (1993) at $Re=100,400, 1000$ and $2000$.}
    \label{fig:cfdf1}
\end{figure}

The computed velocity profiles along the vertical centerline of the cavity for $\mathrm{Re} = 100, 400, 1000$ and $2000$ show good agreement with the benchmark data of Cortes and Miller~\cite{Cortes1993}, as illustrated in Fig.~\ref{fig:cfdf1}. Consequently, the CFD results obtained from this setup are considered reliable and are subsequently used as the ground truth for validating the PINNs predictions.

\subsection{Physics-Informed Neural Networks (PINNs)}

\subsubsection{Neural Network Architecture}

\begin{figure*}[htbp]
    \centering
    \includegraphics[width=1.0\textwidth]{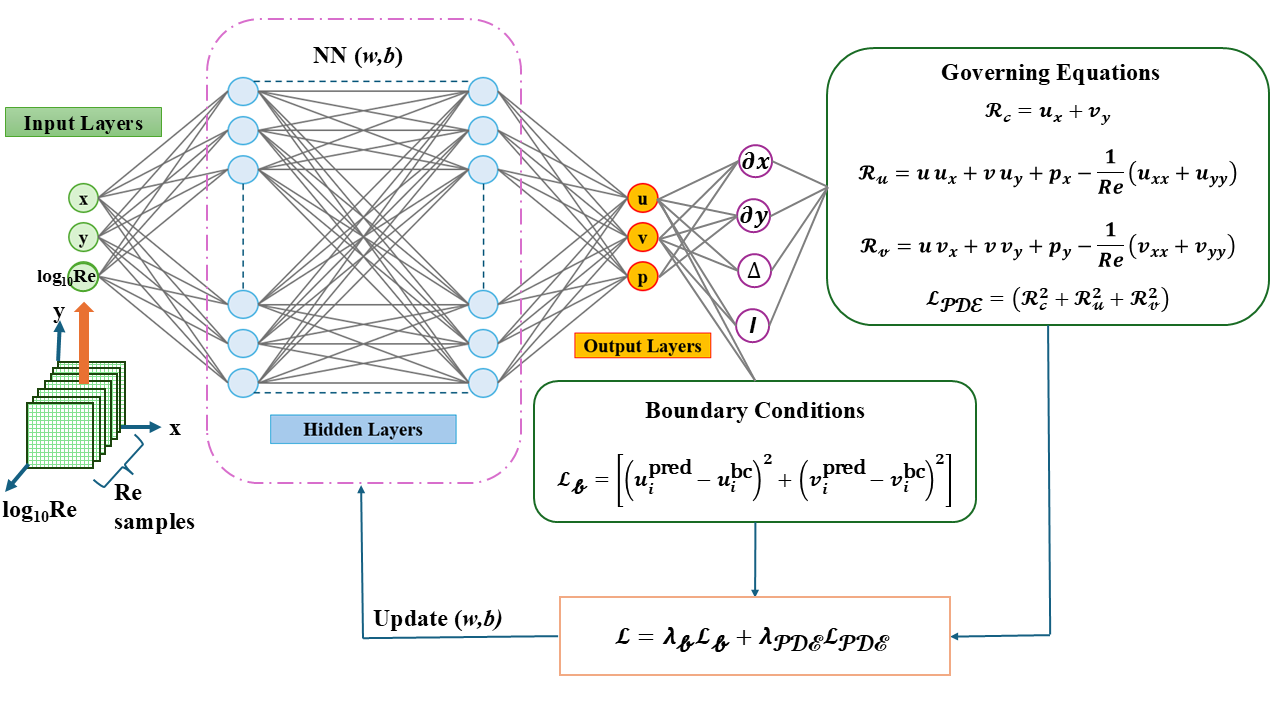}
    \caption{Schematic of the fully connected feed-forward neural network architecture used in this study.}
    \label{fig:nn_architecture}
\end{figure*}

A fully connected feed-forward neural network (FNN) is employed to approximate the velocity and pressure fields within the cavity, as schematically illustrated in Fig.~\ref{fig:nn_architecture}. The network takes three input features: the spatial coordinates $(x,y)$ and the Reynolds number $(\mathrm{Re})$. The network outputs the two velocity components $(u,v)$ and the pressure field $p$.  Treating the Reynolds number as an explicit input parameter enables parametric generalization, allowing the network to learn a continuous solution manifold spanning multiple flow regimes. As shown in the figure, this parametric formulation allows a single neural network to represent solutions corresponding to different Reynolds numbers. Such parameterized PINN approaches have been shown to be effective for solving families of partial differential equations with shared physical structure and for enabling predictive capability beyond fixed boundary conditions~\cite{Sun2020,Meng2020,Clements2025}.

To ensure numerical stability and efficient learning, the Reynolds number input is transformed using a logarithmic mapping and subsequently normalized, as indicated in the input layer of Fig.~\ref{fig:nn_architecture}. This logarithmic transformation compresses the wide dynamic range of Reynolds numbers and prevents it from dominating the spatial inputs. Moreover, this choice is particularly well suited to the hyperbolic tangent ($\tanh$) activation function employed in the hidden layers. Since $\tanh$ exhibits its highest sensitivity and gradient magnitude near zero and rapidly saturates for large input values, centering the log-transformed Reynolds number around zero ensures that the activations remain within the linear regime of $\tanh$, thereby mitigating vanishing-gradient issues and improving convergence during training.

To investigate the influence of network depth and width on predictive accuracy, a parametric study was conducted by varying the number of hidden layers ($H$) and neurons per layer ($N$). Table~\ref{tab:ablation} presents the corresponding mean squared error (MSE), coefficient of determination ($R^2$), and computational time for different network configurations. The results demonstrate that shallow or narrow networks exhibit comparatively larger errors and lower $R^2$ values due to insufficient representational capacity for capturing nonlinear flow physics. Increasing the number of layers and neurons improves prediction accuracy by enhancing the network’s ability to approximate complex flow features. However, excessively increasing the network width beyond the optimal configuration does not necessarily improve accuracy and may lead to over-parameterization and additional computational cost. Among the tested configurations, the architecture with ten hidden layers and eighty neurons per layer $(H=10, N=80)$ achieved the best overall balance between accuracy and computational efficiency, yielding the lowest MSE and highest $R^2$ values for both Reynolds numbers considered. Therefore, this configuration was selected for all subsequent simulations.

 Hyperbolic tangent ($\tanh$) activation functions are employed due to their smoothness, bounded output, and suitability for representing continuous physical fields governed by partial differential equations~\cite{LeCun2012,Raissi2019}. The symmetric output range of $\tanh$ further promotes balanced gradient flow and improved training stability, which is particularly beneficial in physics-informed learning frameworks~\cite{Jagtap2020}.

\begin{table*}[htbp]
\centering
\caption{Influence of network depth and width on prediction accuracy and computational cost for the velocity component $u$.}
\label{tab:ablation}
\begin{tabular}{c c c c c c c c c}
\hline

 & $(H,N)$ & (8,40) & (8,64) & (8,80) & (10,40) & (10,64) & (10,80) & (10,100) \\
\hline
\multicolumn{2}{c}{Time (s) for each 10k epochs} 
& 661 & 717 & 737 & 819 & 824 & 843 & 928 \\
\hline
\multirow{2}{*}{Re = 100} 
& MSE ($u$) 
& $9.59\times10^{-3}$ 
& $7.18\times10^{-3}$ 
& $7.02\times10^{-3}$ 
& $7.99\times10^{-3}$ 
& $4.35\times10^{-3}$ 
& $\mathbf{3.10\times10^{-3}}$ 
& $8.64\times10^{-3}$ \\

& $R^2$ ($u$) 
& 0.808611 
& 0.856709 
& 0.859858 
& 0.840470 
& 0.913217 
& \textbf{0.938099} 
& 0.827421 \\
\hline
\multirow{2}{*}{Re = 200} 
& MSE ($u$) 
& $1.86\times10^{-2}$ 
& $1.59\times10^{-2}$ 
& $1.58\times10^{-2}$ 
& $1.75\times10^{-2}$ 
& $1.13\times10^{-2}$ 
& $\mathbf{9.86\times10^{-3}}$ 
& $1.79\times10^{-2}$ \\

& $R^2$ ($u$) 
& 0.625640 
& 0.681667 
& 0.683605 
& 0.649452 
& 0.772293 
& \textbf{0.801977} 
& 0.641451 \\
\hline
\end{tabular}
\end{table*}

To train the parametric PINNs across a range of flow regimes, Reynolds numbers are sampled within the training interval using a logarithmic-uniform distribution, as illustrated in the Reynolds-number sampling block of Fig.~\ref{fig:nn_architecture}. Logarithmic-uniform sampling ensures balanced coverage across the Reynolds-number space and provides uniform representation in log-space, which is consistent with the logarithmic input transformation. This strategy prevents oversampling at high Reynolds numbers and enhances the network’s ability to learn smooth parametric dependencies on $\mathrm{Re}$.

Network weights are initialized using the Glorot (Xavier) initialization scheme to preserve the variance of activations and gradients across layers, thereby mitigating vanishing or exploding gradient issues and facilitating stable convergence during training~\cite{Glorot2010}. For the data-free PINNs configuration, the training process illustrated in Fig.~\ref{fig:nn_architecture} relies exclusively on physics-based constraints.  During backpropagation, gradients of this total loss with respect to the network parameters $(\mathbf{w}, \mathbf{b})$ are computed, and the weights are updated using gradient-based optimization. Through this update, the network progressively adjusts its parameters so that the predicted fields simultaneously satisfy the governing physical laws and boundary conditions.

This process constitutes one training cycle. Repeating this cycle over many iterations enables the PINNs to minimize the residuals of the Navier-Stokes equations across the sampled Reynolds-number space. As a result, the trained network learns a parametric solution manifold that enforces the underlying physics while generalizing smoothly across different Reynolds numbers within the training range. The selected network depth and width provide sufficient expressive capacity to capture the nonlinear flow features present in the lid-driven cavity flow across the considered range of Reynolds numbers.

\subsubsection{Computational Domain}

\begin{figure}[h]
    \centering
    \includegraphics[width=0.85\linewidth]{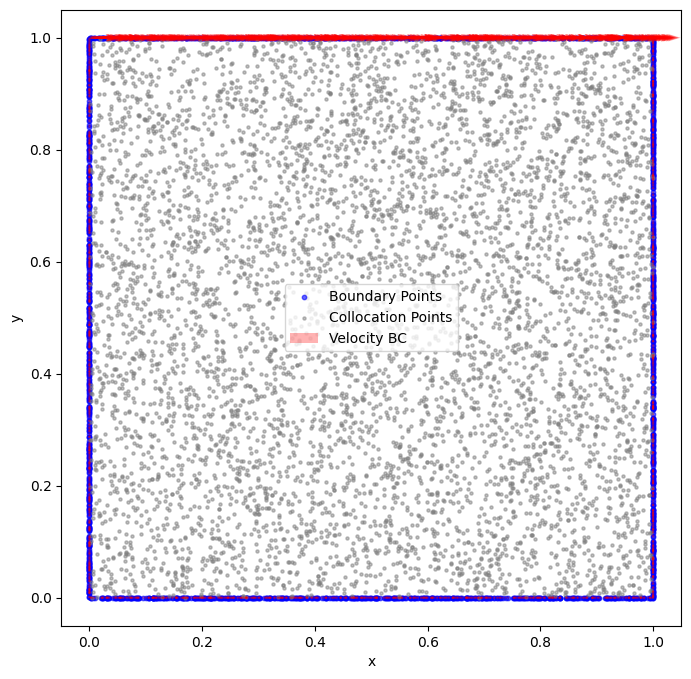}
    \caption{Randomly distributed collocation ($N_{f}$) and boundary ($N_{b}$) points within the $1 \times 1$ cavity domain.}
    \label{fig:collocation}
\end{figure}

The computational domain consists of a two-dimensional square cavity of size 1 × 1. Within this domain, $N_f$ interior collocation points and $N_b$ boundary points are distributed, as illustrated in Fig.~\ref{fig:collocation}. Several sampling strategies were evaluated, including Latin Hypercube sampling, Sobol low-discrepancy sequences, uniform grid sampling, and Monte Carlo sampling. Based on comparative testing, Monte Carlo sampling was selected for the final implementation due to its favorable balance between accuracy, robustness, and computational efficiency.

The collocation points are sampled independently from a uniform distribution over the domain $\Omega = [x_{\min},x_{\max}] \times [y_{\min},y_{\max}]$, such that
\begin{equation}
(x_i,y_i) \sim \mathcal{U}(\Omega)
\end{equation}
The physics-informed residual loss can be interpreted as an integral over the computational domain,
\begin{equation}
\mathcal{L} = \int_{\Omega} \| R(\mathbf{x}) \|^2 \, \mathrm{d}\mathbf{x}
\end{equation}
where $R(\mathbf{x})$ denotes the governing equation residuals. In practice, this integral is approximated using a Monte Carlo estimator,
\begin{equation}
\widehat{\mathcal{L}} = \frac{1}{N_f} \sum_{i=1}^{N_f} \| R(\mathbf{x}_i) \|^2
\end{equation}
which provides an unbiased estimate with a convergence rate of $\mathcal{O}(N_f^{-1/2})$, independent of dimensionality.

Compared to structured grids, random sampling mitigates aliasing effects and grid-induced artifacts, particularly when the residual field contains high-frequency components. The stochasticity introduced by Monte Carlo sampling also benefits gradient-based optimization by improving exploration of the loss landscape~\cite{Raissi2019,Karniadakis2021}. During training, the Adam optimizer is first employed to provide stochastic updates that enable rapid exploration and escape from shallow local minima~\cite{Kingma2015}. Subsequently, the L-BFGS optimizer is applied for deterministic fine-tuning, exploiting the smoother loss surface achieved after Adam pre-training~\cite{Raissi2019,Nocedal2006}. The combined Monte Carlo sampling and Adam-L-BFGS optimization strategy enhances convergence stability and accuracy, particularly for nonlinear, multiscale flows governed by competing convective and diffusive effects~\cite{Wang2021,Jagtap2020}.

\subsubsection{Hyperparameters}

Training of the physics-informed neural network is performed in two stages. In the first stage, the Adam optimizer is used to obtain a coarse approximation of the solution, providing suitable initial weights and biases for the network~\cite{Kingma2015}. In the second stage, the L-BFGS optimizer is employed to further refine convergence and reduce the residuals of the governing equations~\cite{Nocedal2006,Raissi2019}. An adaptive learning rate strategy is adopted, starting from $10^{-3}$ and gradually decreasing to $10^{-6}$ to ensure stable convergence and prevent premature stagnation.

The number of hidden layers ($H$) and neurons per layer ($N$) are selected based on the parametric study summarized in Table~\ref{tab:ablation}, which evaluates the influence of network depth and width on prediction accuracy. A hyperbolic tangent activation function with Glorot initialization is employed in all hidden layers, while the output layer uses a linear activation function to directly represent the physical variables. This architecture provides sufficient expressive capability for accurately learning nonlinear flow fields within the PINNs framework.

\subsubsection{Loss Function Design}

The total loss function, $\mathcal{L}$, is defined as a weighted sum of the boundary-condition loss and the PDE residual loss:

\begin{equation}
\mathcal{L} = \lambda_{\text{b}} \, \mathcal{L}_{\text{b}} 
            + \lambda_{\text{PDE}} \, \mathcal{L}_{\text{PDE}}
\label{eq:loss_total}
\end{equation}

where $\lambda_{\text{b}}$ and $\lambda_{\text{PDE}}$ denote the weighting coefficients associated with the boundary-condition and PDE residual losses, respectively. In the present work, a fixed-weight strategy was employed for all simulations. Equal weighting was assigned to each loss contribution, such that $\lambda_{\text{b}} = \lambda_{\text{PDE}} = 1$.

The PDE loss enforces the incompressible Navier–Stokes equations and continuity. 
The residuals of the governing equations are given by

\begin{align}
\mathcal{R}_{c} &= 
u_{x} + v_{y} 
\label{eq:continuity_residual} \\[6pt]
\mathcal{R}_{u} &= 
u\,u_{x} + v\,u_{y} + p_{x}
- \frac{1}{Re}\left( u_{xx} + u_{yy} \right)
\label{eq:momentum_u_residual} \\[6pt]
\mathcal{R}_{v} &= 
u\,v_{x} + v\,v_{y} + p_{y}
- \frac{1}{Re}\left( v_{xx} + v_{yy} \right)
\label{eq:momentum_v_residual}
\end{align}

where subscripts denote partial derivatives and \(Re\) is the Reynolds number. The PDE loss is defined as the mean squared error (MSE) of these residuals over the collocation points:

\begin{equation}
\mathcal{L}_{\text{PDE}} = 
\frac{1}{N_{f}} \sum_{i=1}^{N_{f}} 
\left( \mathcal{{R}}_{c}^{2} + \mathcal{R}_{u}^{2} + \mathcal{R}_{v}^{2} \right)
\label{eq:loss_pde}
\end{equation}

where $N_{f}$ is the number of collocation points. The PDE residuals were computed using automatic differentiation, thereby embedding the governing physical laws directly into the training process.   

The boundary condition loss enforces the no-slip and lid-driven conditions at the cavity walls:  

\begin{equation}
\mathcal{L}_{\text{b}} =
\frac{1}{N_{b}} \sum_{i=1}^{N_{b}} 
\left[ \left(u^{\text{pred}}_{i} - u^{\text{bc}}_{i}\right)^{2} + 
       \left(v^{\text{pred}}_{i} - v^{\text{bc}}_{i}\right)^{2} \right]
\label{eq:loss_bc}
\end{equation}

where $u^{\text{pred}}_{i}$ and $v^{\text{pred}}_{i}$ are the predicted velocities at boundary point $i$, $u^{\text{bc}}_{i}$ and $v^{\text{bc}}_{i}$ are the prescribed boundary values, and $N_{b}$ is the number of boundary points.

The PINNs framework embeds the governing Navier-Stokes equations and boundary conditions directly into the loss function by penalizing the residuals at collocation points. This formulation follows the original PINNs methodology introduced by Raissi \textit{et al.}~\cite{Raissi2019} and later reviewed comprehensively by Karniadakis \textit{et al.}~\cite{Karniadakis2021}.

This formulation, normalized by the number of collocation and boundary points, ensures that the PINNs solution simultaneously satisfies the governing equations and the prescribed boundary conditions. Several studies have examined the convergence behavior and theoretical properties of PINNs, identifying challenges associated with complex optimization landscapes and limited network expressivity~\cite{Shin2020CICP,Mishra2022}.
\subsubsection{Gradient Norm Analysis}

During the training of PINNs, monitoring the gradient norm is a useful way to evaluate the stability and balance of the optimization process. The gradient norm measures the overall size of the gradients of each loss term with respect to the trainable parameters of the neural network. Tracking this quantity helps detect common training issues such as vanishing gradients, exploding gradients, or imbalance between different loss components.

Imbalanced gradient norms can slow down or prevent convergence in PINNs, especially for convection-dominated problems~\cite{Wang2021,Wang2022loss}. For each loss component $\mathcal{L}_i$, such as the PDE residual loss or the boundary condition (BC) loss, the gradient with respect to the network parameters $\boldsymbol{\theta}$ is denoted by $\nabla_{\boldsymbol{\theta}}\mathcal{L}_i$. The corresponding $\ell_2$-norm of this gradient, referred to as the \emph{gradient norm}, is defined as
\begin{equation}
    \|g\| = \|\nabla_{\boldsymbol{\theta}} \mathcal{L}_i\|_2 =
    \sqrt{\sum_{j=1}^{N_{\theta}} \left( \frac{\partial \mathcal{L}_i}{\partial \theta_j} \right)^2}
    \label{eq:grad_norm_method}
\end{equation}
where $N_{\theta}$ is the total number of trainable parameters in the neural network.

At each training iteration, the gradient norms of the different loss components are computed using automatic differentiation. Comparing these values allows us to assess the relative contribution of each physical constraint to the parameter updates. If one loss term has a much larger gradient norm than the others, it can dominate the training process and lead to an unbalanced solution. In such cases, loss reweighting or adaptive training strategies can be used to restore balance among the different loss terms.

The evolution of the gradient norms during training is recorded and analyzed to ensure that the optimization remains stable and avoids vanishing or exploding gradients. This analysis provides insight into the training dynamics and helps guide the selection of learning rates, loss weights, and optimizer settings in the PINNs framework.

\section{RESULTS AND DISCUSSION}

\subsection{Low Reynolds Number (data-free PINNs)}
In this section, the performance of the proposed PINNs framework is assessed for low-Reynolds-number lid-driven cavity flows, with Reynolds numbers up to $\mathrm{Re} = 300$. It is emphasized that the PINNs model is trained without the use of any CFD data. Instead, learning is driven solely by the governing physics, enforced through the incompressible Navier-Stokes equations and the associated boundary conditions.

Because no data-driven loss terms or external solution samples are incorporated during training, the present approach is referred to as \emph{data-free PINNs}. This setting provides a stringent test of the model’s ability to recover physically consistent flow solutions based exclusively on the embedded physical constraints.

To assess the accuracy and physical fidelity of the data-free PINNs predictions, reference solutions are generated using the open-source finite-volume CFD solver \texttt{OpenFOAM~12}.
These high-quality CFD results are used as the reference solution for comparison. They allow a clear evaluation of how well the model predicts the velocity and pressure fields at different Reynolds numbers. By comparing the PINNs results with the benchmark CFD solutions, we can clearly show how effective the physics-informed learning approach is at capturing the main flow features without using any training data.

\subsubsection{Effect of Sampling Strategies on PINNs Performance}
\label{subsec:sampling_methods}

To evaluate the influence of collocation-point sampling strategies on the performance of PINNs, four different methods were examined: Latin Hypercube sampling, Monte Carlo sampling, Sobol low-discrepancy sampling, and a uniform (structured) grid. Each sampling method was trained over the considered range of Reynolds numbers and then tested for the lid-driven cavity flow at \(Re = 100\). To ensure a fair comparison, the same neural network architecture, identical hyperparameters, and the same number of collocation and boundary points were used for all sampling methods. The convergence behavior, computational cost, and prediction accuracy were compared, and the key results are summarized in Table~\ref{tab:sampling_comparison}.

Among the sampling strategies examined, the Sobol sequence exhibited the fastest convergence to the prescribed residual tolerance of $10^{-3}$, requiring approximately 37{,}000 training epochs. Latin hypercube sampling (LHS) also demonstrated satisfactory convergence behavior, albeit at a higher computational cost. In contrast, the uniform grid showed the slowest convergence and failed to reach the target residual level within 150{,}000 epochs.

Prediction accuracy was measured using the percentage error between the PINNs-predicted and CFD benchmark $u$-velocity along the vertical centerline ($x = 0.5$). Using this metric, Latin Hypercube Sampling (LHS) achieved the highest accuracy, with a mean error of 1.01\% and a maximum local error of 2.67\%. Monte Carlo (MC) sampling gave moderate accuracy, with a mean error of 2.79\%, while requiring far fewer training epochs than both the LHS and uniform grid methods. Although Sobol sampling converged the fastest, it produced the largest errors, with a mean error of 19.31\%.

The lower accuracy of Sobol-based PINNs, despite its fast convergence, is due to the uniformity and low-discrepancy properties of Sobol points. These properties help minimize the global residual quickly but may provide insufficient resolution in areas with sharp gradients, such as near-wall boundary layers and corner vortices in the lid-driven cavity. As a result, the network converges to a solution that satisfies the governing equations in an overall sense but cannot capture fine local flow features accurately. This demonstrates that fast residual convergence does not always guarantee high prediction accuracy in PINNs.

Considering both accuracy and computational cost, Monte Carlo sampling provides the best balance. It achieves reasonable prediction accuracy while significantly reducing training time and computational effort. These advantages make it suitable for repeated PINNs training, parametric studies, and hyperparameter tuning. Therefore, Monte Carlo sampling is chosen as the preferred strategy for the rest of this study.

\setlength{\tabcolsep}{8pt} 

\begin{table*}[htbp]
\centering
\caption{Comparison of sampling methods for PINNs training in the lid-driven cavity at \(Re = 100\). Mean and maximum errors represent the percentage error in the predicted centerline \(u\)-velocity compared to CFD results.}
\label{tab:sampling_comparison}

\setlength{\tabcolsep}{10pt} 

\begin{tabular}{lccccc}
\hline
\textbf{Method} & 
\makecell{\textbf{Epochs to reach}\\\textbf{\(10^{-2}\) residual}} &
\makecell{\textbf{Epochs to reach}\\\textbf{\(10^{-3}\) residual}} &
\makecell{\textbf{Time per}\\\textbf{1000 epochs (s)}} &
\makecell{\textbf{Mean}\\\textbf{Error (\%)}} &
\makecell{\textbf{Maximum}\\\textbf{Error (\%)}} \\
\hline
Latin Hypercube & 35000    & 120000     & 87.0 & 1.01  & 2.67 \\
Monte Carlo     & 18000    & 62000      & 87.5 & 2.79  & 4.56 \\
Sobol           & 16000    & 37000      & 88.5 & 19.31 & 29.63 \\
Uniform Grid    & 83000    & $>150000$  & 86.5 & 5.20  & 8.51 \\
\hline
\end{tabular}
\end{table*}

\subsubsection{Loss Function Convergence}

\begin{figure}[htbp]
    \centering
    \includegraphics[width=1.0\linewidth]{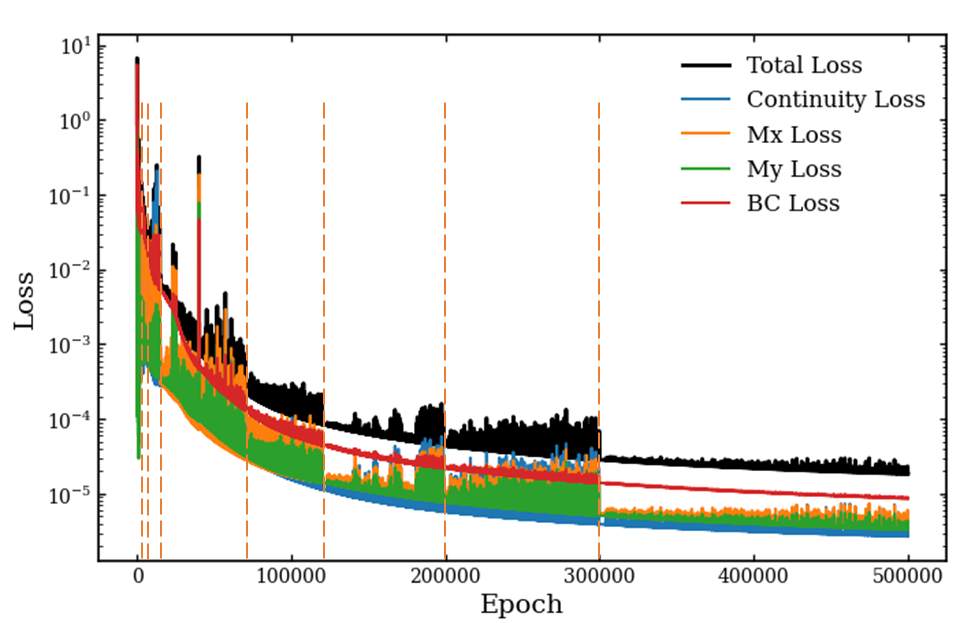}
    \caption{Convergence history of the total loss and its individual components for the lid-driven cavity case. The dotted orange vertical lines indicate the epochs at which the learning rate was adjusted according to the schedule listed in Table~\ref{tab:lr_schedule}.}

    \label{fig:loss_convergence}
\end{figure}

The convergence behavior of the total loss and its individual components is illustrated in Fig.~\ref{fig:loss_convergence}. The total loss function consists of contributions from the governing equations (PDE loss) and the boundary conditions (BC loss), with the PDE loss further decomposed into the continuity loss and the momentum losses in the $x$- and $y$-directions, corresponding to the residuals of the Navier-Stokes equations. As shown in Fig.~\ref{fig:loss_convergence}, the boundary loss 
exhibits a relatively higher magnitude compared to the PDE components during training, reflecting the stricter enforcement of boundary constraints. To ensure stable and balanced convergence, a piecewise-constant learning rate schedule was employed. The training started with an initial learning rate of $5\times10^{-3}$, which was gradually reduced at selected epochs, as summarized in Table~\ref{tab:lr_schedule}. A larger learning rate in the early stages facilitates rapid learning, while progressively smaller learning rates allow finer parameter updates as the solution converges. The dotted orange vertical lines in Fig.~\ref{fig:loss_convergence} mark the epochs at which the learning rate was adjusted, in accordance with the schedule listed in Table~\ref{tab:lr_schedule}. 

The model was trained for a total of 500{,}000 epochs using the Adam optimizer during the initial phase to achieve rapid convergence and stable descent of the loss function, followed by the L-BFGS optimizer for high-precision fine-tuning of the network parameters after the Adam training stage. The gradual and smooth reduction of all loss components demonstrates the stable and physically consistent convergence behavior of the proposed PINNs framework for the lid-driven cavity flow case.

\setlength{\tabcolsep}{10pt}
\begin{table}[htbp]
    \centering
    \caption{A piecewise-constant learning rate schedule was used during PINNs training.}
    \label{tab:lr_schedule}
    \begin{tabular}{c c}
        \hline
        \textbf{Epoch Range} & \textbf{Learning Rate} \\
        \hline
        $0$ -- $1{,}000$           & $5\times10^{-3}$ \\
        $1{,}001$ -- $5{,}000$     & $1\times10^{-3}$ \\
        $5{,}001$ -- $15{,}000$    & $5\times10^{-4}$ \\
        $15{,}001$ -- $70{,}000$   & $1\times10^{-4}$ \\
        $70{,}001$ -- $120{,}000$  & $5\times10^{-5}$ \\
        $120{,}001$ -- $200{,}000$ & $1\times10^{-5}$ \\
        $200{,}001$ -- $300{,}000$ & $5\times10^{-6}$ \\
        $>300{,}000$               & $1\times10^{-6}$ \\
        \hline
    \end{tabular}
\end{table}

\subsubsection{Gradient Norm Plot}
\begin{figure}[htbp]
    \centering
    \includegraphics[width=1.0\linewidth]{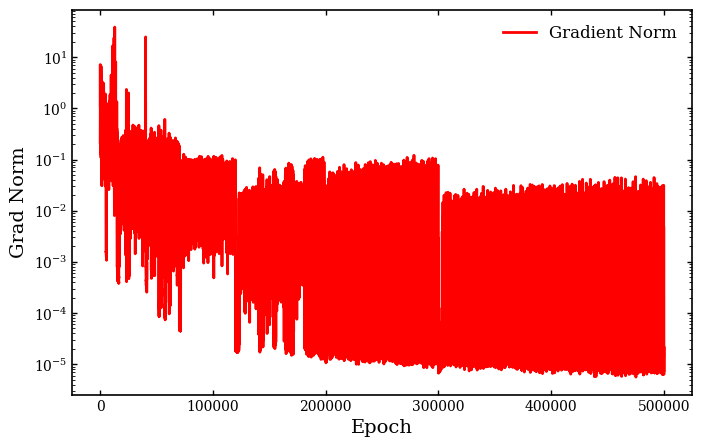}
    \caption{Evolution of gradient norms during training for data-free PINNs at different Reynolds numbers.}
    \label{fig:grad_norm}
\end{figure}

To assess training stability and diagnose stiffness in the optimization process, gradient norm analysis is employed by monitoring the magnitude of gradients associated with different loss components. 
Figure~\ref{fig:grad_norm} illustrates the variation of the gradient norm with respect to the number of training epochs. The gradient norm provides insight into the stability of the optimization process and the learning capability of the network. At the beginning of the training, the gradient norm lies within the healthy range of $10^{-3}$ to $10^{1}$, indicating an adequate magnitude of gradients for effective parameter updates. This range ensures that the network learns efficiently without experiencing gradient explosion or vanishing.

During training, the gradient norm gradually decreases, showing that the optimizer is moving toward a local minimum of the loss. However, after about 150{,}000 epochs, the gradient norm drops below $10^{-4}$, leaving the healthy range. This indicates the onset of the gradient vanishing problem, where the parameter updates become extremely small. As a result, the training slows down and eventually stagnates, which is reflected by the plateau in the total loss shown in Fig.~\ref{fig:loss_convergence}. At this point, the loss function is almost flat, and further training does not significantly improve the model's accuracy.

The observed trend suggests that extending the training beyond this point provides diminishing returns while increasing computational cost. Therefore, it is recommended to terminate the training after a few hundred thousand epochs once the gradient norm approaches the lower bound of the healthy region. Monitoring the gradient norm thus serves as a practical criterion for early stopping, preventing unnecessary computations and ensuring efficient use of computational resources.

\subsubsection{Contours of $u$, $v$, $p$, and Vorticity}

The predicted velocity, pressure, and vorticity contours provide a comprehensive assessment of the capability of the proposed PINNs framework to reconstruct the full flow field within the lid-driven cavity domain. In contrast to localized centerline profile comparisons, these contour-based evaluations enable examination of the global spatial accuracy of the model and its ability to capture derived flow quantities associated with rotational flow structures. As the Reynolds number increases, the flow develops a dominant primary vortex along with progressively stronger secondary corner vortices, which are consistent with classical numerical and experimental studies of lid-driven cavity flow~\cite{Ghia1982,Botella1998,Erturk2006,Shankar2000}. These characteristics introduce significant optimization challenges for purely physics-based PINNs without supervised data assistance.

\begin{figure*}[htbp]
    \centering
    \includegraphics[scale=0.55]{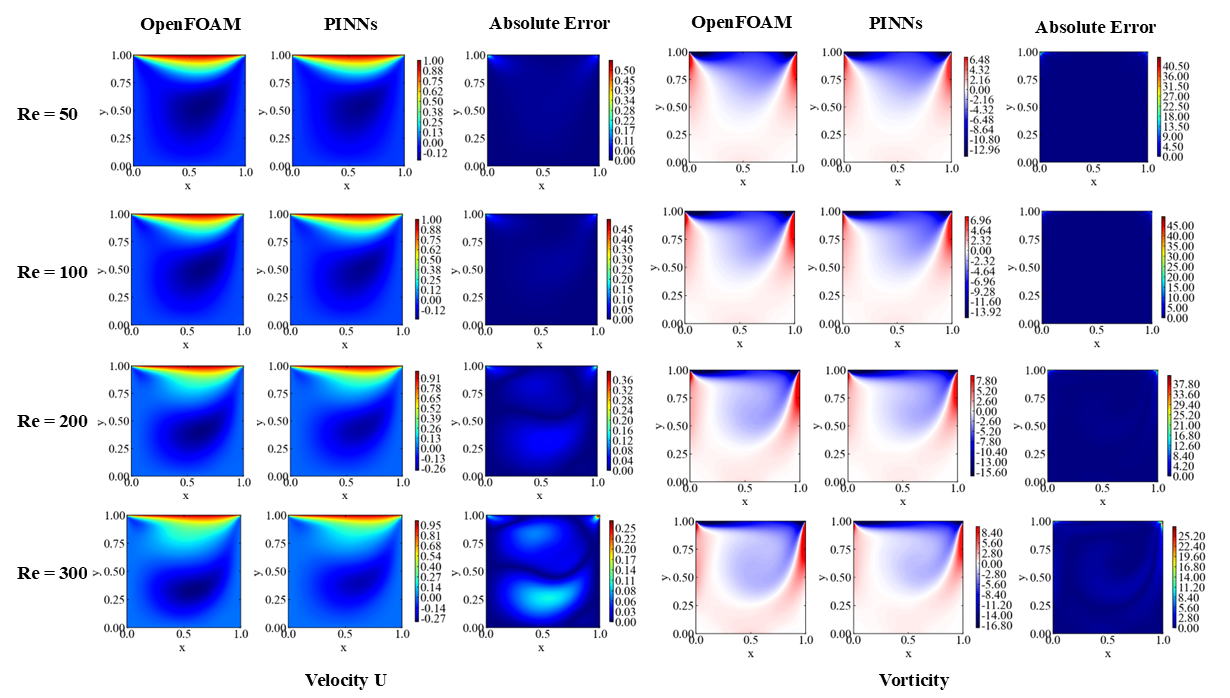}
    \caption{Comparison of horizontal velocity ($u$) and vorticity contours between OpenFOAM (ground truth) and PINNs for lid-driven cavity flow at $\mathrm{Re}=50,\ 100,\ 200,$ and $300$. The corresponding absolute error distributions are also shown. Increasing discrepancies at higher Reynolds numbers highlight the difficulty of purely physics-based PINNs in accurately resolving convection-dominated flow structures.}
    \label{fig:fig6}
\end{figure*}

\begin{figure*}[htbp]
    \centering
    \includegraphics[scale=0.66]{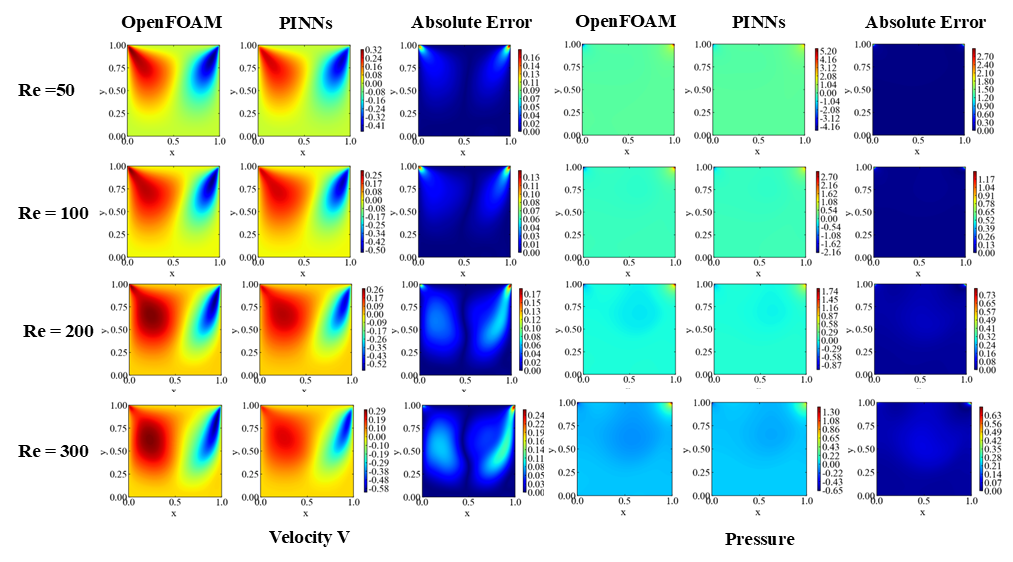}
    \caption{Comparison of vertical velocity ($v$) and pressure ($p$) contours between OpenFOAM (ground truth) and PINNs for lid-driven cavity flow at $\mathrm{Re}=50,\ 100,\ 200,$ and $300$. The absolute error contours show increasing discrepancies in the $v$-velocity field for $\mathrm{Re}\geq200$, particularly near shear layers and recirculation regions where convection effects dominate.}
    \label{fig:fig7}
\end{figure*}

Contour comparisons for Reynolds numbers $\mathrm{Re}=50,\ 100,\ 200,$ and $300$ are presented in Figures~\ref{fig:fig6} and~\ref{fig:fig7}. Figure~\ref{fig:fig6} presents the horizontal velocity component ($u$) together with the corresponding vorticity contours. Each row compares the OpenFOAM reference solution, the PINNs prediction, and the associated absolute error distribution. At lower Reynolds numbers ($\mathrm{Re}=50$ and $100$), the PINNs predictions remain in close agreement with the CFD solutions, accurately reproducing the primary vortex structure and the overall velocity distribution throughout the cavity.

However, as the Reynolds number increases to $\mathrm{Re}=200$ and $300$, noticeable discrepancies begin to emerge due to the increasingly convection-dominated nature of the flow. In particular, the prediction errors become more pronounced near the vortex core, shear-layer regions, and moving-lid corners where strong gradients develop. At $\mathrm{Re}=300$, the degradation becomes more evident in both the horizontal velocity and vorticity fields, indicating the limitations of purely data-free PINNs in accurately resolving high-gradient flow structures at higher Reynolds numbers.

The vorticity contours provide a particularly stringent assessment because vorticity depends directly on spatial velocity gradients. The increasing discrepancies observed in the vorticity field at higher Reynolds numbers further demonstrate that purely physics-based PINNs struggle to fully capture convection-dominated flow physics without additional supervision. These observations motivate the introduction of sparse CFD data supervision and transfer learning in the subsequent high-Reynolds-number hybrid PINNs framework.

Figure~\ref{fig:fig7} presents the vertical velocity component ($v$) together with the pressure field ($p$). Similar trends are observed for these quantities. While strong agreement is obtained at lower Reynolds numbers, the prediction accuracy of the $v$-velocity field deteriorates noticeably for $\mathrm{Re}\geq200$. The largest discrepancies occur near the upper cavity corners and along the recirculation regions, where convection effects dominate and the secondary flow structures become increasingly sensitive to local velocity gradients. In contrast, the pressure field remains comparatively smoother and is reconstructed more accurately across all Reynolds numbers, although localized deviations still appear near high-gradient regions at $\mathrm{Re}=300$.

Overall, the contour comparisons confirm that the proposed PINNs framework accurately reconstructs the global flow field for low-to-moderate Reynolds numbers. However, as convection effects become dominant at higher Reynolds numbers, the accuracy of the purely data-free PINNs deteriorates, particularly for the $v$-velocity and vorticity fields, highlighting the need for hybrid data-assisted strategies in convection-dominated regimes.

\subsubsection{Velocity Profiles}
Velocity profiles along the cavity centerlines are compared with CFD reference solutions following standard validation practices commonly adopted in lid-driven cavity benchmarks~\cite{Ghia1982,Botella1998}. In addition to the conventional centerline comparisons, velocity profiles at multiple vertical and horizontal sampling locations throughout the cavity domain are also examined to provide a more comprehensive assessment of the full-field predictive accuracy of the proposed PINNs framework.

\begin{figure}[htbp]
    \centering
    \includegraphics[width=0.50\textwidth]{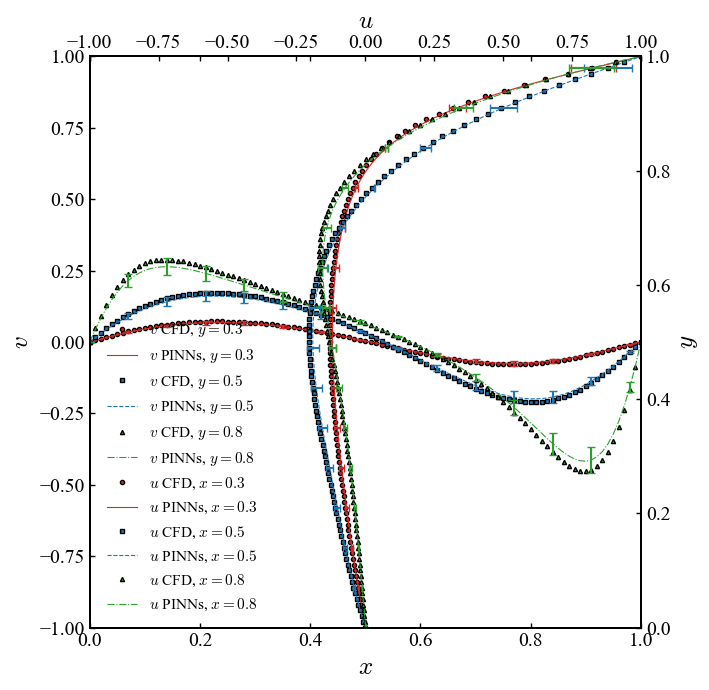}
    \caption{Comparison of $u$-velocity profiles along vertical lines ($x = 0.3$, $0.5$, $0.8$) and $v$-velocity profiles along horizontal lines ($y = 0.3$, $0.5$, $0.8$) between PINNs predictions and OpenFOAM results at $Re = 50$. Error bars represent $\pm 10\%$ of the CFD ground truth.}
    \label{fig:combined_velocity_profiles05}
\end{figure}

\begin{figure}[htbp]
    \centering
    \includegraphics[width=0.50\textwidth]{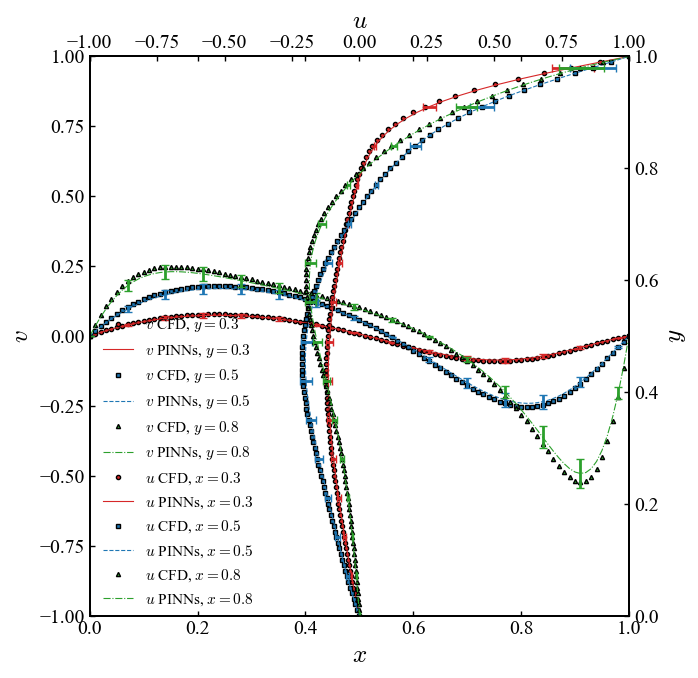}
    \caption{Comparison of $u$-velocity profiles along vertical lines ($x = 0.3$, $0.5$, $0.8$) and $v$-velocity profiles along horizontal lines ($y = 0.3$, $0.5$, $0.8$) between PINNs predictions and OpenFOAM results at $Re = 100$. Error bars represent $\pm 10\%$ of the CFD ground truth.}
    \label{fig:combined_velocity_profiles1}
\end{figure}

\begin{figure}[htbp]
    \centering
    \includegraphics[width=0.50\textwidth]{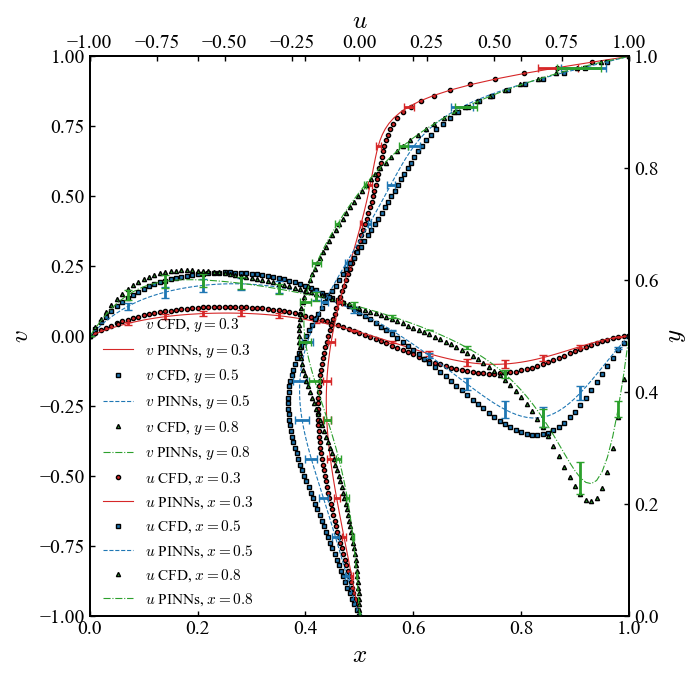}
    \caption{Comparison of $u$-velocity profiles along vertical lines ($x = 0.3$, $0.5$, $0.8$) and $v$-velocity profiles along horizontal lines ($y = 0.3$, $0.5$, $0.8$) between PINNs predictions and OpenFOAM results at $Re = 200$. Error bars represent $\pm 10\%$ of the CFD ground truth.}
    \label{fig:combined_velocity_profiles2}
\end{figure}

Figures~\ref{fig:combined_velocity_profiles05}--\ref{fig:combined_velocity_profiles2} present quantitative comparisons between the PINNs predictions and the CFD reference solutions obtained using \texttt{OpenFOAM~12} for lid-driven cavity flows at $Re = 50$, $100$, and $200$, respectively. The figures show the horizontal velocity component ($u$) evaluated along selected vertical lines at $x = 0.3$, $0.5$, and $0.8$, together with the vertical velocity component ($v$) evaluated along selected horizontal lines at $y = 0.3$, $0.5$, and $0.8$. Across all Reynolds numbers and sampled locations, the PINNs predictions exhibit strong agreement with the CFD results, accurately capturing the velocity distributions, flow gradients, and cavity flow symmetry. The error bars represent $\pm 10\%$ of the CFD ground truth for $Re = 50$, $100$, and $200$. The predicted velocity profiles remain within these acceptable deviation limits, demonstrating the robustness and reliability of the proposed physics-informed framework in reproducing low- and moderate-Reynolds-number cavity flows.

\subsubsection{Integral Force Analysis on the Moving Lid}

In addition to the local velocity profile comparisons, the predictive capability of the data-free PINNs framework was further assessed using integral force quantities computed on the moving upper wall of the cavity. Unlike pointwise comparisons, integral force measures provide a more rigorous global assessment of the solution quality because they directly depend on the spatial gradients of the velocity and pressure fields over the entire wall boundary.

The total force acting on the moving lid consists of contributions from viscous shear stress and pressure forces. The tangential shear force was evaluated from the wall shear stress distribution, while the normal force component was computed using the mean-corrected pressure field to eliminate the gauge-pressure ambiguity inherent in incompressible flow simulations. The resultant integral force therefore provides an additional validation metric for evaluating the physical consistency of the PINNs predictions relative to the \texttt{OpenFOAM~12} reference solutions.

Table~\ref{tab:pure_force_comparison} presents the comparison of the tangential force ($F_x$), pressure force ($F_y$), and total resultant force ($F_{\mathrm{total}}$) between the data-free PINNs predictions and the corresponding CFD solutions for Reynolds numbers ranging from $\mathrm{Re}=50$ to $300$. The percentage errors associated with each force component are also reported.

The results demonstrate that the data-free PINNs framework predicts the viscous shear force with good accuracy across the investigated Reynolds number range, with errors generally remaining below approximately $11\%$. The total resultant force also shows strong agreement with the CFD reference solutions, indicating that the PINNs model successfully captures the dominant momentum transport mechanisms within the cavity flow. Although comparatively larger discrepancies are observed in the pressure-force component, the magnitude of the pressure contribution remains significantly smaller than the viscous contribution. Consequently, the overall resultant force predictions remain physically consistent and in close agreement with the OpenFOAM solutions.

\begin{table*}[htbp]
\centering
\caption{Comparison of integral force components on the moving lid between data-free PINNs and OpenFOAM solutions for the lid-driven cavity flow.}
\label{tab:pure_force_comparison}
\begin{tabular}{c c c c c c c c c c}
\hline
$\mathrm{Re}$ &
$F_{x,\mathrm{CFD}}$ &
$F_{x,\mathrm{PINNs}}$ &
Error (\%) &
$F_{y,\mathrm{CFD}}$ &
$F_{y,\mathrm{PINNs}}$ &
Error (\%) &
$F_{\mathrm{total,CFD}}$ &
$F_{\mathrm{total,PINNs}}$ &
Error (\%) \\
\hline

50  & 0.3294364 & 0.2920563 & 11.35 & 0.0047281 & 0.0030989 & 34.46 & 0.3294703 & 0.2920727 & 11.35 \\

80  & 0.2114689 & 0.1923692 & 9.03 & 0.0047543 & 0.0042000 & 11.66 & 0.2115223 & 0.1924151 & 9.03 \\

100 & 0.1725990 & 0.1589752 & 7.89 & 0.0047847 & 0.0041439 & 13.39 & 0.1726653 & 0.1590292 & 7.90 \\

150 & 0.1209959 & 0.1121401 & 7.32 & 0.0048775 & 0.0037644 & 22.82 & 0.1210942 & 0.1122033 & 7.34 \\

200 & 0.0950754 & 0.0891298 & 6.25 & 0.0049502 & 0.0035043 & 29.21 & 0.0952042 & 0.0891986 & 6.31 \\

300 & 0.0686443 & 0.0625402 & 8.89 & 0.0050222 & 0.0025497 & 49.23 & 0.0688277 & 0.0625921 & 9.06 \\

\hline
\end{tabular}
\end{table*}

\subsubsection{Influence of Collocation-Point Density on Model Accuracy}

To investigate the influence of collocation-point density on the predictive capability of the PINNs framework, three spatial discretizations were considered: a coarse grid (3{,}300 points), a medium grid (5{,}200 points), and a fine grid (11{,}700 points). Each configuration included both interior collocation points enforcing the governing equations and boundary points enforcing the Dirichlet conditions.

The prediction accuracy was quantified using the mean squared error (MSE) and the coefficient of determination ($R^2$), which are commonly employed performance metrics for physics-informed learning frameworks~\cite{Hafezianzadeh2023PINN,Mamud2024PINNmetrics}. Lower MSE values indicate smaller discrepancies between the PINNs predictions and the CFD reference solutions, while $R^{2}$ values approaching unity indicate stronger predictive agreement.

To further examine the effect of collocation density under increasingly convection-dominated conditions, the grid-sensitivity study was extended up to $Re = 300$. In addition, the computational cost associated with each grid density was evaluated using the training time per 10{,}000 epochs.

Table~\ref{tab:grid_convergence} summarizes the resulting MSE and $R^{2}$ values for the $u$-velocity field across Reynolds numbers ranging from $Re = 50$ to $300$, together with the corresponding computational cost.

The results indicate that PINNs predictions exhibit relatively low sensitivity to grid refinement at lower Reynolds numbers ($Re \leq 100$). Even the coarse discretization provides reasonably accurate solutions with comparatively high $R^{2}$ values, indicating that the dominant laminar flow physics can be captured using a moderate number of collocation points.

As the Reynolds number increases, stronger nonlinear convection and sharper velocity gradients lead to increasing prediction errors for all grid resolutions. Although finer collocation distributions generally improve accuracy, particularly for $Re = 150$-$300$, the improvements become progressively limited at higher Reynolds numbers. For example, at $Re = 300$, refining the grid from 5{,}200 to 11{,}700 points improves the MSE from $9.82\times10^{-2}$ to $6.12\times10^{-2}$ and increases $R^2$ from $0.7817$ to $0.8617$, but the computational cost more than doubles.

These observations indicate that increasing collocation density alone does not guarantee substantial accuracy improvements in convection-dominated regimes ($Re > 200$). In such cases, optimization stiffness, loss imbalance between convective and diffusive terms, and the spectral bias of PINNs may become increasingly dominant factors limiting predictive performance.

The computational-cost analysis further highlights the trade-off between accuracy and efficiency. The coarse, medium, and fine grids required approximately 205 s, 262 s, and 540 s per 10{,}000 epochs, respectively. Although the fine grid consistently produced the most accurate predictions, it also incurred significantly higher computational expense.

\setlength{\tabcolsep}{8pt}
\begin{table*}[htbp]
    \centering
    \caption{Effect of collocation-point density on PINNs accuracy and computational cost. Mean squared error (MSE), coefficient of determination ($R^{2}$), and computational time per 10,000 epochs for coarse (3{,}300 points), medium (5{,}200 points), and fine (11{,}700 points) grids across different Reynolds numbers.}
    \label{tab:grid_convergence}
    \begin{tabular}{c|ccc|ccc}
        \hline
        & \multicolumn{3}{c|}{\textbf{MSE}} 
        & \multicolumn{3}{c}{\boldmath$R^{2}$} \\
        
        \textbf{Re} 
        & Coarse & Medium & Fine 
        & Coarse & Medium & Fine \\
        \hline
        
        50  & $1.78\times10^{-4}$ & $1.88\times10^{-4}$ & $1.96\times10^{-4}$ 
            & 0.9860 & 0.9832 & 0.9846 \\

        80  & $8.37\times10^{-4}$ & $6.38\times10^{-4}$ & $1.98\times10^{-4}$ 
            & 0.9740 & 0.9863 & 0.9939 \\

        100 & $1.77\times10^{-3}$ & $1.19\times10^{-3}$ & $2.17\times10^{-4}$ 
            & 0.9647 & 0.9768 & 0.9937 \\

        150 & $7.07\times10^{-3}$ & $5.19\times10^{-3}$ & $1.93\times10^{-3}$ 
            & 0.9571 & 0.9534 & 0.9827 \\

        200 & $2.42\times10^{-2}$ & $1.18\times10^{-2}$ & $1.11\times10^{-2}$ 
            & 0.8784 & 0.9093 & 0.9442 \\

        250 & $5.99\times10^{-2}$ & $4.76\times10^{-2}$ & $2.73\times10^{-2}$ 
            & 0.8080 & 0.8475 & 0.9032 \\

        300 & $1.10\times10^{-1}$ & $9.82\times10^{-2}$ & $6.12\times10^{-2}$ 
            & 0.7544 & 0.7817 & 0.8617 \\
            
        \hline
        \textbf{Time per 10k epochs (s)} 
            & 205 & 262 & 540 
            & -- & -- & -- \\
        \hline
    \end{tabular}
\end{table*}

\begin{figure*}[htbp]
    \centering
    \includegraphics[width=0.45\textwidth]{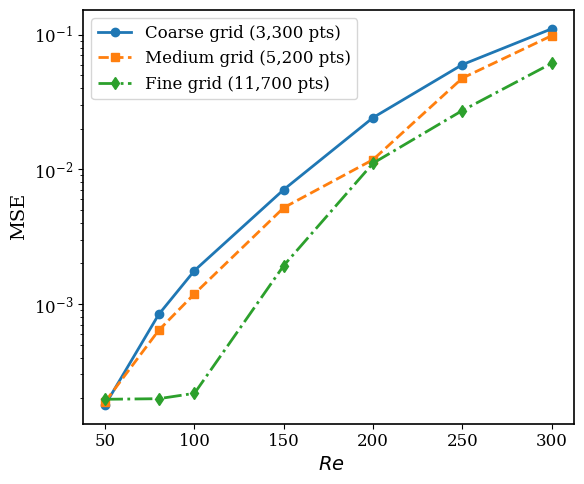}
    \includegraphics[width=0.45\textwidth]{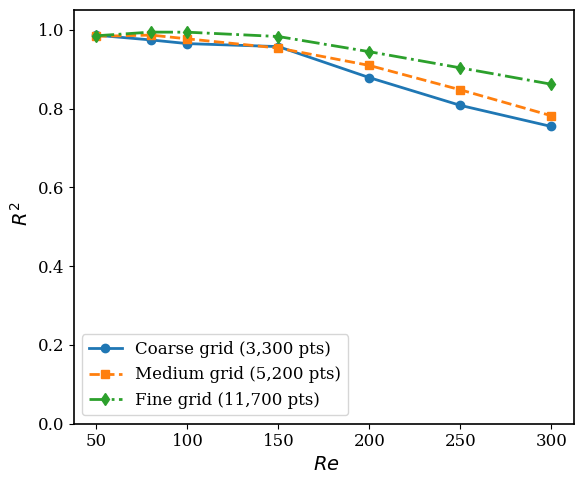}
    \caption{
        Influence of collocation-point density on PINNs accuracy for the lid-driven cavity problem. 
        Left: Mean squared error (MSE) of the predicted $u$ velocity fields for coarse (3{,}300 points), 
        medium (5{,}200 points), and fine (11{,}700 points) grids. 
        Right: Coefficient of determination ($R^2$) for the same grids. 
    }
    \label{fig:collocation_density}
\end{figure*}

Figure~\ref{fig:collocation_density} illustrates the influence of collocation-point density on the predictive accuracy of the PINNs framework for the lid-driven cavity problem.

The left panel presents the MSE variation of the $u$-velocity field with Reynolds number for the three grid resolutions. The fine grid generally provides lower prediction errors, particularly at intermediate Reynolds numbers. However, beyond $Re \approx 200$, the reduction in MSE obtained through additional collocation points becomes comparatively moderate.

The right panel presents the corresponding $R^2$ values. At lower Reynolds numbers, all three grids maintain relatively high $R^2$ values, indicating that the essential flow physics are captured accurately even with coarse discretizations. At higher Reynolds numbers, the fine grid improves the overall prediction fidelity, although increasing collocation density alone does not fully prevent the degradation in accuracy caused by increasingly convection-dominated flow behavior.

In summary, the results demonstrate that moderate collocation densities are sufficient for obtaining accurate predictions at low and intermediate Reynolds numbers. While finer collocation distributions improve solution fidelity to some extent, the present study indicates that collocation refinement alone is insufficient to fully overcome the optimization challenges associated with convection-dominated high-Reynolds-number flows.

\subsubsection{Model Accuracy: Comparison Between CFD and PINNs Predictions}

To assess both the interpolation and extrapolation capabilities of the proposed PINNs framework, the model predictions were compared against high-fidelity CFD solutions at multiple Reynolds numbers. The Reynolds number range used for model training, $50 \le \mathrm{Re} \le 300$, is defined as the \emph{interpolation region}, while Reynolds numbers outside this interval are treated as the \emph{extrapolation region}. This distinction enables a systematic evaluation of the model’s ability to generalize within and beyond the trained Reynolds-number space.

Table~\ref{tab:mse_r2_resultsP} summarizes the MSE and $R^2$ values for the velocity components $(u,v)$ and pressure field $p$ across both interpolation and extrapolation regimes. Within the interpolation region ($50 \le \mathrm{Re} \le 300$), the PINNs framework demonstrates good predictive capability, with relatively low MSE values and consistently high $R^2$ values for the velocity fields. The strongest agreement is observed at lower and moderate Reynolds numbers, where viscous diffusion remains sufficiently dominant and the governing equations are comparatively easier for PINNs to optimize.

As the Reynolds number increases, a gradual deterioration in predictive accuracy is observed, particularly beyond $Re \approx 200$. This behavior is associated with the increasingly convection-dominated nature of the flow, where sharper velocity gradients, stronger nonlinear transport, and thin shear layers emerge. Under such conditions, standard data-free PINNs experience increasing optimization difficulties due to spectral bias and loss imbalance between convective and diffusive terms. Consequently, both the MSE values increase and the corresponding $R^2$ values decrease progressively with Reynolds number, even within the interpolation regime.

The pressure field exhibits comparatively larger MSE values, which is expected since pressure in incompressible flows is defined only up to an arbitrary constant and generally exhibits weaker spatial variations than velocity. Nevertheless, after mean correction, the PINNs framework is still able to recover the overall pressure distribution reasonably well within the trained Reynolds-number range.

\begin{figure*}[htbp]
    \centering
    \includegraphics[width=0.48\textwidth]{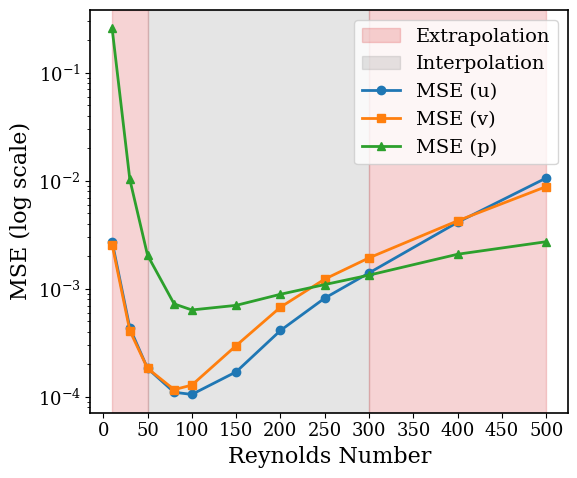}
    \includegraphics[width=0.48\textwidth]{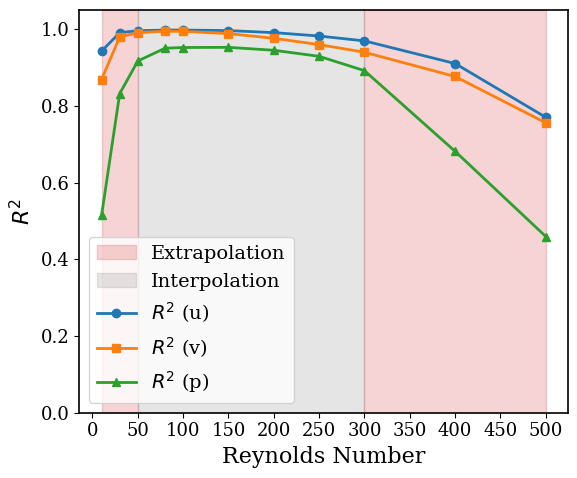}
    \caption{
        Quantitative comparison of PINNs predictions with CFD data across different Reynolds numbers. The gray and red regions denote the interpolation and extrapolation regimes, respectively.
        Left: MSE of the predicted velocity components $u$, $v$, and pressure $p$.
        Right: $R^{2}$ comparing PINNs predictions with CFD reference solutions.
    }
    \label{fig:mse_r2_combinedP}
\end{figure*}

Figure~\ref{fig:mse_r2_combinedP} summarizes these trends visually. The gray shaded region represents the interpolation region corresponding to the training range ($50 \le \mathrm{Re} \le 300$), while the red shaded region denotes the extrapolation regime. The left panel presents the variation of MSE with Reynolds number, showing progressively increasing prediction errors as the flow becomes more convection dominated. The right panel shows the corresponding $R^2$ values, which remain high at low and moderate Reynolds numbers but decrease gradually at higher Reynolds numbers.

Outside the interpolation region (e.g., $\mathrm{Re}=10, 30$ and $\mathrm{Re}\ge400$), the degradation in predictive accuracy becomes more pronounced. In particular, the extrapolation results at higher Reynolds numbers indicate the limitations of purely physics-informed training without additional supervision. These observations motivated the introduction of sparse CFD-assisted supervision and transfer learning strategies in the subsequent high-Reynolds-number hybrid PINNs framework, where limited supervised data are incorporated to stabilize training and improve predictive accuracy in strongly convection-dominated regimes.

Overall, the results demonstrate that the proposed data-free PINNs framework achieves good predictive performance for low and moderate Reynolds-number cavity flows, while also highlighting the increasing challenges encountered as convection progressively dominates diffusion at higher Reynolds numbers.

\begin{table*}[ht]
\centering
\caption{MSE and $R^2$ results for different Reynolds numbers. 
Rows shaded in gray indicate the interpolation region ($50 \le \mathrm{Re} \le 300$), 
while unshaded rows correspond to the extrapolation region.}
\label{tab:mse_r2_resultsP}
\begin{tabular}{c|ccc|ccc}
\hline
 & \multicolumn{3}{c|}{\textbf{MSE}} & \multicolumn{3}{c}{\boldmath$R^{2}$} \\
\textbf{Re} & $u$ & $v$ & $p$ & $u$ & $v$ & $p$ \\
\hline
10  & $2.74\times10^{-3}$ & $2.57\times10^{-3}$ & $2.58\times10^{-1}$ & 0.9427 & 0.8687 & 0.5145 \\
30  & $4.36\times10^{-4}$ & $4.10\times10^{-4}$ & $1.03\times10^{-2}$ & 0.9908 & 0.9793 & 0.8320 \\

\rowcolor{gray!15}
50  & $1.84\times10^{-4}$ & $1.84\times10^{-4}$ & $2.05\times10^{-3}$ & 0.9961 & 0.9910 & 0.9170 \\
\rowcolor{gray!15}
80  & $1.11\times10^{-4}$ & $1.17\times10^{-4}$ & $7.27\times10^{-4}$ & 0.9976 & 0.9946 & 0.9504 \\
\rowcolor{gray!15}
100 & $1.06\times10^{-4}$ & $1.29\times10^{-4}$ & $6.39\times10^{-4}$ & 0.9977 & 0.9943 & 0.9522 \\
\rowcolor{gray!15}
150 & $1.71\times10^{-4}$ & $2.99\times10^{-4}$ & $7.05\times10^{-4}$ & 0.9963 & 0.9884 & 0.9526 \\
\rowcolor{gray!15}
200 & $4.12\times10^{-4}$ & $6.78\times10^{-4}$ & $8.92\times10^{-4}$ & 0.9910 & 0.9762 & 0.9450 \\
\rowcolor{gray!15}
250 & $8.22\times10^{-4}$ & $1.23\times10^{-3}$ & $1.10\times10^{-3}$ & 0.9821 & 0.9597 & 0.9292 \\
\rowcolor{gray!15}
300 & $1.41\times10^{-3}$ & $1.94\times10^{-3}$ & $1.34\times10^{-3}$ & 0.9695 & 0.9398 & 0.8918 \\

400 & $4.14\times10^{-3}$ & $4.27\times10^{-3}$ & $2.09\times10^{-3}$ & 0.9104 & 0.8766 & 0.6815 \\
500 & $1.06\times10^{-2}$ & $8.85\times10^{-3}$ & $2.73\times10^{-3}$ & 0.7702 & 0.7558 & 0.4594 \\
\hline
\end{tabular}
\end{table*}

\subsection{Generalizability Assessment on Backward-Facing Step Flow}

To further investigate the regime-dependent predictive capability of parameterized PINNs beyond the lid-driven cavity benchmark, an additional validation study was conducted using the two-dimensional backward-facing step (BFS) flow. In contrast to the closed recirculating cavity configuration, the BFS problem involves flow separation, shear-layer development, recirculation, and downstream reattachment, thereby providing a substantially different flow topology for assessing the robustness of physics-informed learning across incompressible flow regimes.

The backward-facing step is a classical benchmark problem in incompressible fluid mechanics and has been extensively used for validating numerical methods and turbulence models due to its sensitivity to Reynolds number and its characteristic separated-flow behavior \cite{armaly1983experimental,driver1985features,kaiktsis1996structure}. In the context of the present study, the BFS configuration serves primarily as a secondary validation case for examining whether the regime-dependent behavior observed for lid-driven cavity flow also persists for a qualitatively different separated-flow problem.

\subsubsection{Problem Description and Flow Physics}

At low Reynolds numbers, the backward-facing step flow remains steady and laminar, with a primary recirculation zone forming immediately downstream of the step. As the Reynolds number increases, the separated shear layer becomes progressively stronger and the reattachment point moves farther downstream. The increasing dominance of convective transport also introduces sharper velocity gradients and stronger shear-layer interactions, which increase the optimization difficulty for pure physics-informed learning approaches.

The present study considers steady laminar flow conditions over the Reynolds number range $50 \le \mathrm{Re} \le 300$ for training of the data-free parameterized PINNs model. Unlike the later high-Reynolds-number lid-driven cavity investigations, no sparse CFD supervision is introduced for the BFS case. The objective here is specifically to assess the capability and limitations data-free parameterized PINNs under physics-only training for a different incompressible flow topology.

\begin{figure*}[htbp]
\centering
\includegraphics[width=1\linewidth]{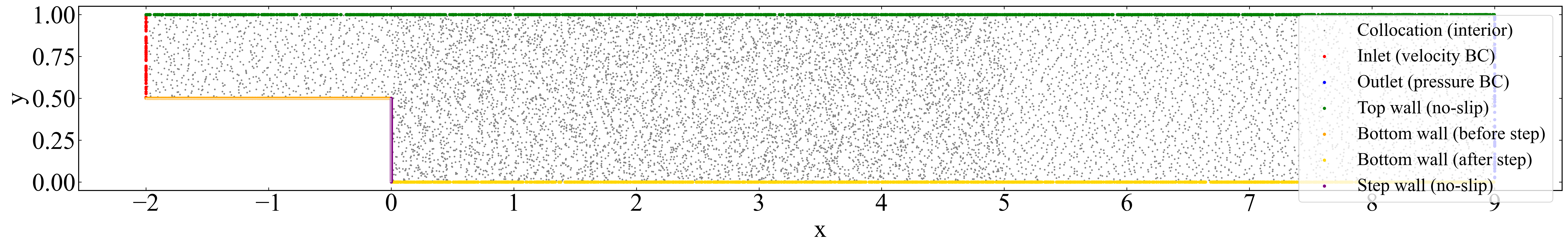}
\caption{Schematic representation of the two-dimensional backward-facing step (BFS) computational domain used in the present study.}
\label{fig:bfs_geometry}
\end{figure*}

The computational geometry for the backward-facing step is illustrated schematically in Fig.~\ref{fig:bfs_geometry}. The domain consists of an inlet section of height \(H_i\), followed by a sudden expansion to the channel height \(H_c\) and the step height \(h = H_c - H_i\). The geometric parameters used in the present study are summarized below:

\[
L_{\mathrm{in}} = 2.0,
\qquad
L_{\mathrm{out}} = 9.0,
\qquad
H_i = 0.5,
\qquad
H_c = 1.0
\]

where \(L_{\mathrm{in}}\) and \(L_{\mathrm{out}}\) denote the inlet and outlet lengths, respectively. The inlet length was selected sufficiently large to avoid entrance-development effects prior to the step, while the outlet length was chosen long enough to ensure that the downstream boundary condition does not influence the reattachment region.

Since accurate prediction of the separated shear layer and downstream reattachment location is central to the BFS problem, additional collocation points were distributed in the downstream region up to \(x = 5 \) in order to improve resolution of the reattachment dynamics and near-wall recovery region.

High-fidelity CFD simulations were performed using \texttt{OpenFOAM~12} over the Reynolds number range $50 \le \mathrm{Re} \le 500$ to generate benchmark reference solutions for comparison with the PINNs predictions.

\subsubsection{Reattachment Length Validation}

The computed reattachment lengths were validated against the experimental measurements of Armaly et al.~\cite{armaly1983experimental}, which remain one of the standard benchmark references for laminar backward-facing step flow.

The reattachment length \(x_1/h\), defined as the downstream location where the wall shear stress changes sign and the separated flow reattaches to the lower wall, was used as the principal validation metric \cite{armaly1983experimental,driver1985features}.

\begin{figure}[htbp]
\centering
\includegraphics[width=0.95\linewidth]{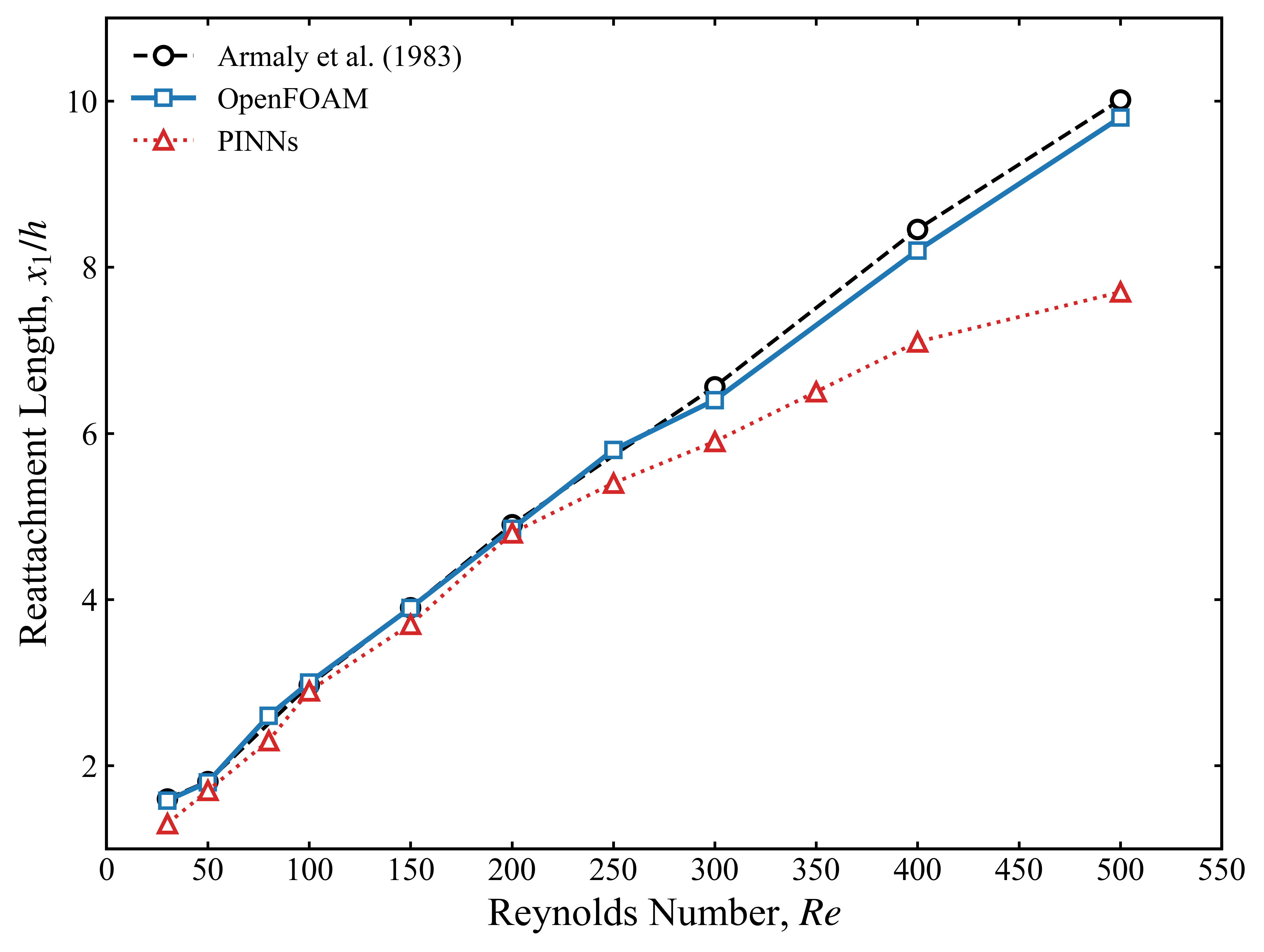}
\caption{Comparison of normalized reattachment length \(x_1/h\) for backward-facing step flow between the experimental data of Armaly (1983), OpenFOAM simulations, and PINNs predictions.}
\label{fig:bfs_reattachment}
\end{figure}

Figure~\ref{fig:bfs_reattachment} and Table~\ref{tab:bfs_reattachment} compare the normalized reattachment lengths obtained from the present \texttt{OpenFOAM~12} simulations, the PINNs predictions, and the experimental measurements of Armaly et al.~\cite{armaly1983experimental} over the Reynolds number range \(30 \le Re \le 500\).

Although the parameterized PINNs model was trained only within the Reynolds number interval \(50 \le Re \le 300\), the reattachment-length predictions were evaluated over a broader Reynolds number range in order to assess both interpolation and extrapolation behavior. It can be observed that the OpenFOAM simulations exhibit excellent agreement with the experimental benchmark data throughout the investigated Reynolds number range.

The parameterized PINNs predictions also capture the overall trend of increasing reattachment length with Reynolds number. In particular, the PINNs predictions up to approximately \(Re = 200\) show excellent agreement with both the OpenFOAM simulations and the experimental measurements of Armaly et al.~\cite{armaly1983experimental}. However, moderate deviations become progressively more noticeable at higher Reynolds numbers, especially beyond the primary training regime, where the separated shear layer becomes increasingly convection dominated and more difficult to approximate using purely physics-based training.

This behavior is consistent with the regime-dependent trends observed previously for the lid-driven cavity flow, namely that data-free PINNs remain highly effective for low-to-moderate Reynolds-number regimes but gradually lose predictive robustness as convective transport becomes dominant. The BFS results therefore provide additional evidence that the limitations of data-free PINNs are closely linked to the underlying flow physics rather than solely to optimization procedures. 
These observations further motivate the sparse-supervised hybrid strategy introduced subsequently for the high-Reynolds-number lid-driven cavity problem, where localized CFD supervision is incorporated only after the degradation of pure physics-informed learning becomes significant.

\begin{table}[htbp]
\centering
\caption{Comparison of normalized reattachment length \(x_1/h\) for backward-facing step flow between the experimental measurements of Armaly et al.~\cite{armaly1983experimental}, OpenFOAM simulations, and PINNs predictions. Although the PINNs model was trained for \(50 \le Re \le 300\), predictions are additionally reported outside the training interval to assess extrapolation behavior.}
\label{tab:bfs_reattachment}
\begin{tabular}{cccc}
\hline
\(Re\) & Armaly et al.~\cite{armaly1983experimental} & OpenFOAM & PINNs \\
\hline
30  & 1.60  & 1.58 & 1.30 \\
50  & 1.81  & 1.80 & 1.70 \\
80  & --    & 2.60 & 2.30 \\
100 & 2.96  & 3.00 & 2.90 \\
150 & 3.90  & 3.90 & 3.70 \\
200 & 4.90  & 4.85 & 4.80 \\
250 & --    & 5.80 & 5.40 \\
300 & 6.56  & 6.40 & 5.90 \\
400 & 8.45  & 8.20 & 7.10 \\
500 & 10.10 & 9.80 & 7.70 \\
\hline
\end{tabular}
\end{table}

\subsubsection{Velocity Profile Comparisons}

To further evaluate the predictive capability and parametric generalization behavior of the data-free PINNs framework, detailed comparisons of the normalized streamwise velocity profiles \((u/U_{\mathrm{ref}})\) were performed at multiple streamwise locations inside the backward-facing step (BFS) domain. The velocity profiles were extracted along vertical lines located at \(x = -1, 0, 1, 2, 3, 4, 5, 6, 7,\) and \(8\), thereby enabling assessment of the model performance in the upstream region, near the separation point, throughout the recirculation zone, and within the downstream flow recovery region.

\begin{figure*}[htbp]
\centering
\includegraphics[width=1\linewidth]{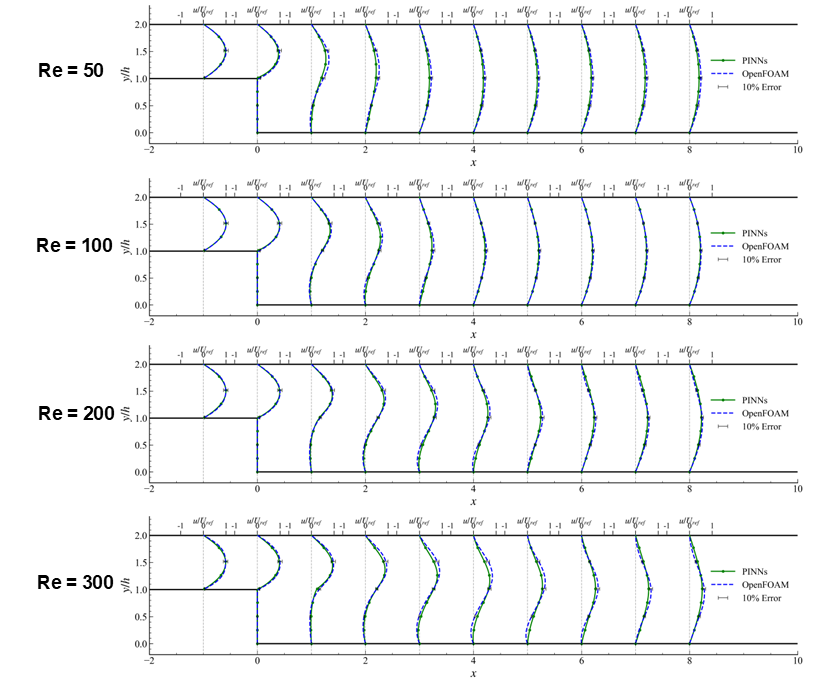}
\caption{Comparison of normalized streamwise velocity profiles \((u/U_{\mathrm{ref}})\) between the data-free parameterized PINNs framework and OpenFOAM simulations at different streamwise locations \((x=-1 \text{ to } 8)\) inside the backward-facing step domain for Reynolds numbers \(Re=50,\ 100,\ 200,\) and \(300\).}
\label{fig:bfs_velocity_profiles}
\end{figure*}

Figure~\ref{fig:bfs_velocity_profiles} presents the comparison between the PINNs predictions and the corresponding \texttt{OpenFOAM~12} reference solutions for \(Re = 50,\ 100,\ 200,\) and \(300\). To quantify the agreement between the two approaches, a \(\pm 10\%\) error limit relative to the \texttt{OpenFOAM~12} solution is incorporated in the form of horizontal error bars.

For low and moderate Reynolds numbers (\(Re \leq 200\)), the predicted velocity profiles remain in close agreement with the CFD reference solutions across most streamwise locations. The PINNs framework successfully captures the primary flow separation near the step, the development of the recirculation region, shear-layer evolution, and downstream velocity recovery. These results demonstrate that data-free parameterized PINNs can generalize effectively for diffusion-dominated and moderately convection-dominated laminar BFS flows without requiring supervised CFD data.

However, as the Reynolds number increases toward \(Re = 300\), noticeable deviations begin to emerge, particularly within the separated shear layer and reattachment region where stronger convective transport and sharper velocity gradients are present. Although the overall flow behavior remains qualitatively captured, the degradation in local profile accuracy indicates increasing optimization difficulty for pure physics-informed learning in convection-dominated regimes.

These observations are consistent with the trends previously observed for the lid-driven cavity flow and further support the central hypothesis of the present study: data-free PINNs perform reliably in low-Reynolds-number regimes but progressively lose accuracy as convection-dominated effects become stronger. The BFS results therefore provide additional evidence that sparse supervised correction strategies may become necessary at higher Reynolds numbers to improve prediction robustness and stabilize training in strongly convection-dominated separated flows.

\subsubsection{Contour Plots of \(u\), \(v\), \((\partial p/\partial x)\), and Vorticity}

To further assess the capability and limitations of the data-free parameterized PINNs framework in capturing separated-flow physics, contour comparisons of the horizontal velocity (\(u\)), vertical velocity (\(v\)), streamwise pressure-gradient field \((\partial p/\partial x)\), and vorticity were performed for representative Reynolds numbers \(Re = 50,\ 100,\ 150,\ 200,\) and \(300\).

\begin{figure*}[htbp]
\centering
\includegraphics[width=1\linewidth]{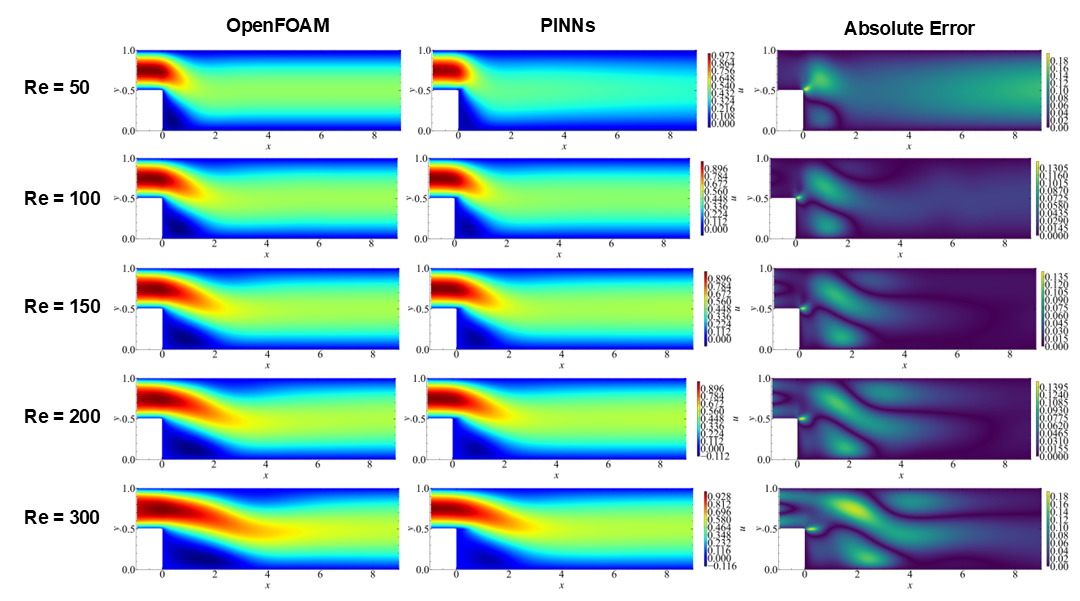}
\caption{Comparison of horizontal velocity (\(u\)) contours between OpenFOAM simulations, PINNs predictions, and corresponding absolute error distributions for backward-facing step flow at \(Re = 50,\ 100,\ 150,\ 200,\) and \(300\).}
\label{fig:bfs_u_contours}
\end{figure*}

\begin{figure*}[htbp]
\centering
\includegraphics[width=1\linewidth]{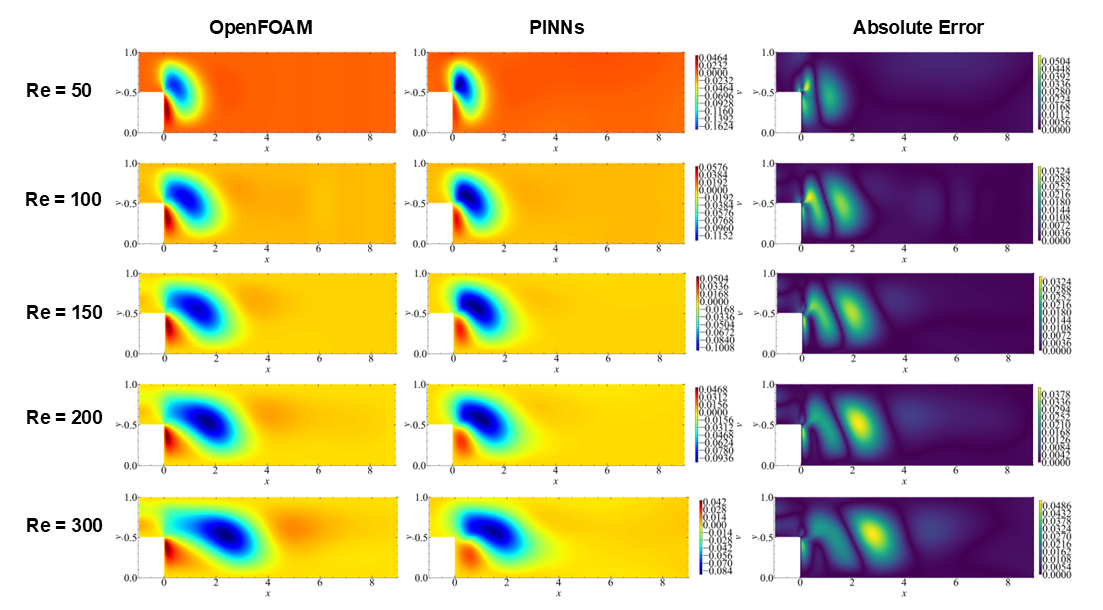}
\caption{Comparison of vertical velocity (\(v\)) contours between OpenFOAM simulations, PINNs predictions, and corresponding absolute error distributions for backward-facing step flow at \(Re = 50,\ 100,\ 150,\ 200,\) and \(300\).}
\label{fig:bfs_v_contours}
\end{figure*}

\begin{figure*}[htbp]
\centering
\includegraphics[width=1\linewidth]{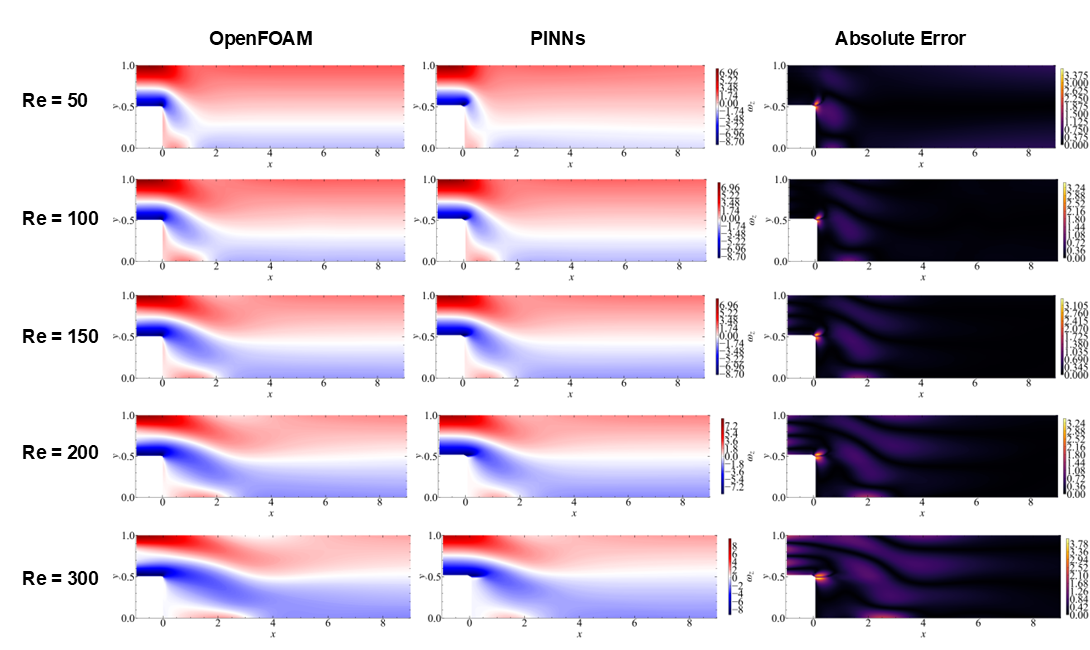}
\caption{Comparison of vorticity contours between OpenFOAM simulations, PINNs predictions, and corresponding absolute error distributions for backward-facing step flow at \(Re = 50,\ 100,\ 150,\ 200,\) and \(300\).}
\label{fig:bfs_vorticity_contours}
\end{figure*}

\begin{figure*}[htbp]
\centering
\includegraphics[width=1\linewidth]{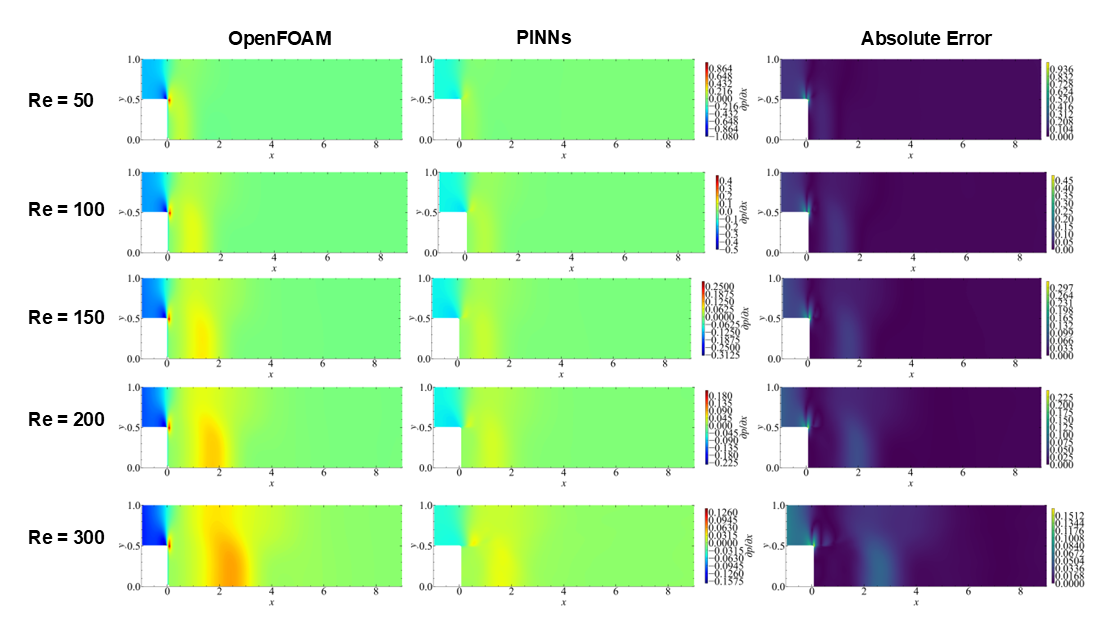}
\caption{Comparison of streamwise pressure-gradient field \((\partial p/\partial x)\) contours between OpenFOAM simulations, PINNs predictions, and corresponding absolute error distributions for backward-facing step flow at \(Re = 50,\ 100,\ 150,\ 200,\) and \(300\).}
\label{fig:bfs_pressure_contours}
\end{figure*}

Figures~\ref{fig:bfs_u_contours}-\ref{fig:bfs_pressure_contours} compare the \texttt{OpenFOAM~12} reference solutions, PINNs predictions, and the corresponding absolute error distributions for the investigated Reynolds numbers. The contour plots provide detailed visualization of the capability of the data-free PINNs model to reconstruct the separated shear layer, recirculation region, and downstream reattachment behavior across different Reynolds-number regimes.

Figure~\ref{fig:bfs_u_contours} presents the contours of the streamwise velocity component \(u\). For low and moderate Reynolds numbers, the PINNs predictions accurately reproduce the primary flow acceleration near the upper wall, the low-velocity recirculation region downstream of the step, and the gradual redevelopment of the channel flow farther downstream. The predicted reattachment location and shear-layer evolution remain in close agreement with the \texttt{OpenFOAM~12} simulations for \(Re \leq 200\). The corresponding absolute error contours indicate that the dominant discrepancies are localized primarily near the separated shear layer and reattachment region where strong gradients are present.

As the Reynolds number increases toward \(Re = 300\), the separated shear layer becomes increasingly convection dominated, resulting in larger localized deviations in the predicted velocity field. Although the global flow topology remains reasonably captured, the error amplification near the reattachment region indicates growing difficulty for pure physics-informed optimization in accurately resolving sharp convective structures without additional supervision.

The contours of the vertical velocity component \(v\), shown in Fig.~\ref{fig:bfs_v_contours}, further demonstrate the ability of the PINNs framework to capture the secondary flow structures generated by the separated shear layer. The model successfully predicts the upward and downward motions associated with recirculation and downstream recovery for lower Reynolds numbers. However, similar to the \(u\)-velocity field, the error magnitude increases progressively with Reynolds number, particularly near the shear layer where the solution becomes increasingly stiff and multiscale.

The vorticity contour comparisons shown in Fig.~\ref{fig:bfs_vorticity_contours} provide additional insight into the rotational flow structures generated downstream of the backward-facing step. Since vorticity depends directly on local velocity gradients, it represents a particularly challenging quantity for data-free PINNs in convection-dominated regimes. The PINNs predictions accurately capture the primary vortex structures and downstream vorticity transport for low Reynolds numbers, while localized discrepancies become more pronounced as the Reynolds number increases. The largest errors are concentrated near the step corner and along the separated shear layer, where the gradients are strongest.

Representative contour comparisons of the streamwise pressure-gradient field \((\partial p/\partial x)\) are provided in Fig.~\ref{fig:bfs_pressure_contours}. The PINNs framework successfully predicts the pressure drop immediately downstream of the step together with the downstream pressure recovery behavior. Nevertheless, increasing deviations are again observed at higher Reynolds numbers, particularly near the reattachment region where pressure gradients become sharper and more strongly coupled to convection-dominated flow dynamics.

In summary, the contour comparisons demonstrate that data-free parameterized PINNs can accurately reconstruct the dominant separated-flow features of backward-facing step flow at low and moderate Reynolds numbers. However, the progressive degradation in localized accuracy at higher Reynolds numbers further supports the regime-dependent behavior identified throughout the present study, namely that data-free PINNs remain effective in diffusion-dominated regimes but become increasingly challenged in convection-dominated flows. These observations motivate the introduction of sparse supervised correction strategies for higher-Reynolds-number flows, as investigated subsequently for the lid-driven cavity configuration.

\subsubsection{Quantitative Accuracy Assessment}

The quantitative accuracy of the data-free parameterized PINNs framework was further evaluated using the MSE and \(R^2\) metrics for the streamwise velocity (\(u\)), transverse velocity (\(v\)), and streamwise pressure-gradient field \((\partial p/\partial x)\).

Table~\ref{tab:bfs_metrics} summarizes the resulting error metrics across the investigated Reynolds number range. Overall, the framework demonstrates strong predictive capability for low and moderate Reynolds-number backward-facing step flows while successfully capturing the dominant separated-flow physics without requiring supervised CFD data.

For the streamwise velocity component \(u\), the PINNs model achieves high prediction accuracy throughout the primary training regime. The \(R^2\) values increase from approximately \(0.72\) at \(Re=30\) to values exceeding \(0.98\) within the Reynolds-number interval \(100 \le Re \le 150\), indicating excellent agreement with the OpenFOAM reference solutions. The corresponding MSE values remain comparatively small, reaching minimum values on the order of \(10^{-4}\).

Similarly, the transverse velocity component \(v\) is predicted with high accuracy for low and moderate Reynolds numbers. The \(R^2\) values remain above \(0.90\) for Reynolds numbers up to approximately \(Re=150\), with relatively small MSE values throughout the training regime. However, as the Reynolds number increases, a systematic degradation in predictive accuracy is observed, particularly for \(Re \geq 200\), where the separated shear layer becomes increasingly convection dominated and more difficult to approximate using pure physics-informed learning alone.

The pressure-gradient predictions exhibit similar trends. Good agreement with the CFD reference solutions is obtained within the low-Reynolds-number regime, while progressively larger deviations emerge at higher Reynolds numbers. Nevertheless, the PINNs framework continues to capture the overall pressure-gradient distribution and downstream recovery characteristics associated with the separated flow.

It is important to note that the best predictive performance is consistently achieved within the low-to-moderate Reynolds-number regime, where viscous effects remain sufficiently strong and the flow field remains relatively smooth. As convection-dominated behavior intensifies at higher Reynolds numbers, optimization stiffness and sharp local gradients increasingly affect the predictive capability of data-free PINNs. The observed reduction in \(R^2\) values and increase in MSE metrics at elevated Reynolds numbers are therefore consistent with the regime-dependent limitations identified throughout the present study.

Taken together, the quantitative analyses further demonstrate that data-free parameterized PINNs generalize effectively for low-Reynolds-number incompressible flows and can successfully capture separated-flow physics in the backward-facing step configuration. However, the progressive degradation in prediction accuracy at higher Reynolds numbers provides additional motivation for incorporating sparse supervised correction strategies in convection-dominated regimes, as explored subsequently for the lid-driven cavity flow.

\begin{table*}[htbp]
\centering
\caption{Quantitative error metrics for backward-facing step flow predictions using the data-free parameterized PINNs framework. Gray-shaded rows indicate the Reynolds number range used during training (\(50 \le Re \le 300\)).}
\label{tab:bfs_metrics}
\begin{tabular}{ccccccc}
\hline
\(Re\) & MSE\(_u\) & \(R^2_u\) & MSE\(_v\) & \(R^2_v\) & MSE\(_{\partial p/\partial x}\) & \(R^2_{\partial p/\partial x}\) \\
\hline

30  & \(1.2310 \times 10^{-2}\) & 0.7215 & \(1.8265 \times 10^{-4}\) & 0.8649 & \(4.9169 \times 10^{-3}\) & 0.8618 \\

40  & \(6.4274 \times 10^{-3}\) & 0.8599 & \(1.0924 \times 10^{-4}\) & 0.9116 & \(2.3104 \times 10^{-3}\) & 0.8905 \\

\rowcolor{gray!15}
50  & \(3.7041 \times 10^{-3}\) & 0.9222 & \(8.5125 \times 10^{-5}\) & 0.9258 & \(1.4635 \times 10^{-3}\) & 0.8972 \\

\rowcolor{gray!15}
80  & \(1.1828 \times 10^{-3}\) & 0.9777 & \(6.3287 \times 10^{-5}\) & 0.9349 & \(7.5273 \times 10^{-4}\) & 0.8857 \\

\rowcolor{gray!15}
100 & \(8.3694 \times 10^{-4}\) & 0.9852 & \(6.5478 \times 10^{-5}\) & 0.9281 & \(5.7873 \times 10^{-4}\) & 0.8778 \\

\rowcolor{gray!15}
150 & \(9.5484 \times 10^{-4}\) & 0.9855 & \(8.0354 \times 10^{-5}\) & 0.8953 & \(4.2652 \times 10^{-4}\) & 0.8461 \\

\rowcolor{gray!15}
200 & \(1.7596 \times 10^{-3}\) & 0.9767 & \(1.3002 \times 10^{-4}\) & 0.8303 & \(4.6010 \times 10^{-4}\) & 0.7712 \\

\rowcolor{gray!15}
250 & \(2.8117 \times 10^{-3}\) & 0.9668 & \(1.5948 \times 10^{-4}\) & 0.7576 & \(5.8052 \times 10^{-4}\) & 0.6319 \\

\rowcolor{gray!15}
300 & \(4.5376 \times 10^{-3}\) & 0.9515 & \(2.2050 \times 10^{-4}\) & 0.6664 & \(7.1633 \times 10^{-4}\) & 0.4617 \\

400 & \(9.1652 \times 10^{-3}\) & 0.9168 & \(3.3801 \times 10^{-4}\) & 0.4810 & \(1.0555 \times 10^{-3}\) & 0.0124 \\

\hline
\end{tabular}
\end{table*}

\subsection{High Reynolds Number Regime (with Transfer Learning)}

In this section, the performance of the parameterized PINNs framework is further examined for higher Reynolds numbers in the lid-driven cavity flow, specifically up to \(\mathrm{Re}=2000\). The previous low-Reynolds-number investigations for both the lid-driven cavity and backward-facing step flows demonstrated that pure physics-informed learning can accurately reconstruct incompressible flow fields in diffusion-dominated regimes without requiring supervised CFD data. However, the results also revealed a systematic degradation in predictive accuracy as the Reynolds number increased and the flow became increasingly convection dominated.

This degradation is associated with the increasing stiffness of the Navier-Stokes equations at high Reynolds numbers, where sharp velocity gradients, thin shear layers, and multiscale flow structures emerge. In such regimes, PINNs are known to suffer from spectral bias and optimization difficulties, which limit their ability to accurately represent high-frequency solution components~\cite{Wang2021}. The backward-facing step results presented earlier further confirmed this regime-dependent behavior, where prediction errors progressively increased near the separated shear layer and reattachment region as the Reynolds number increased.In addition, classical results on the predictability of multi-scale fluid systems indicate that errors originating at small or insufficiently constrained scales can rapidly propagate to larger scales, imposing intrinsic limits on achievable predictive accuracy~\cite{Lorenz1969Predictability}.

At high \(\mathrm{Re}\), the convective terms in the Navier-Stokes equations dominate over the diffusive terms. Consequently, during optimization, the PDE-based residual loss becomes increasingly difficult to balance, often causing the optimizer to converge toward suboptimal regions of the loss landscape. As a result, data-free PINNs may fail to accurately capture steep gradients and localized flow structures characteristic of convection-dominated flows. These observations suggest that physics-only learning becomes insufficient beyond certain Reynolds-number regimes, even though it remains highly effective at lower Reynolds numbers.

Motivated by these observations, sparse supervised CFD data are introduced in the present high-Reynolds-number regime through an additional data-loss term, resulting in a hybrid PINNs framework. Importantly, the objective is not to replace physics-informed learning with dense supervised training, but rather to minimally correct the failure modes of data-free PINNs in convection-dominated regimes. The supervised data therefore act as localized corrective guidance while the governing equations continue to provide the dominant learning constraints.

In addition, transfer learning is employed by initializing the network weights using a pretrained model obtained from lower Reynolds-number regimes, where data-free PINNs already demonstrated strong predictive capability.This strategy leverages previously learned physical representations of the flow and has been shown to accelerate convergence and improve solution quality in PINNs~\cite{Wang2025arxiv,Prantikos2023}. The combined use of transfer learning and localized sparse supervision therefore provides a practical mechanism for bridging low- and high-Reynolds-number regimes within a unified parameterized framework.

After training, the model predictions are compared against ground-truth CFD solutions for Reynolds numbers both within the training range, referred to as the \emph{interpolation region}, and outside the training range, referred to as the \emph{extrapolation region}. This evaluation enables systematic assessment of the effectiveness of sparse supervised correction in improving the robustness of PINNs for convection-dominated incompressible flows while maintaining parametric generalization capability across Reynolds numbers.

\subsubsection{Loss Function Design for High Reynolds Numbers}

For high Reynolds number cases, the total loss function is augmented with an additional data-driven term to compensate for the dominance of the convective terms in the Navier-Stokes equations. The modified loss function is defined as

\begin{equation}
\mathcal{L} =
\lambda_{\text{b}}\,\mathcal{L}_{\text{b}} +
\lambda_{\text{PDE}}\,\mathcal{L}_{\text{PDE}} +
\lambda_{\text{D}}\,\mathcal{L}_{\text{D}}
\label{eq:loss_total_highRe}
\end{equation}

where $\lambda_{\text{b}}$, $\lambda_{\text{PDE}}$, and $\lambda_{\text{D}}$ are weighting coefficients.Equal weighting was assigned to each loss contribution, such that $\lambda_{\text{b}} = \lambda_{\text{PDE}} = \lambda_{\text{D}} = 1$.

The PDE and BC loss retain the same structure as in the data-free PINNs formulation and enforces the continuity and momentum equations along with the boundary conditions.
To improve accuracy at high $Re$, a data loss term is introduced using CFD data from \texttt{OpenFOAM~12}:

\begin{equation}
\mathcal{L}_{\text{D}} =
\frac{1}{N_D}\sum_{i=1}^{N_d}
\left[
 (u^{\text{pred}}_i - u^{\text{D}}_i)^2 +
 (v^{\text{pred}}_i - v^{\text{D}}_i)^2 +
 (p^{\text{pred}}_i - p^{\text{D}}_i)^2
\right]
\label{eq:loss_data}
\end{equation}

where $(u^{\text{D}}_i, v^{\text{D}}_i, p^{\text{D}}_i)$ denote the reference CFD values and $N_D$ is the number of data points.

This hybrid loss formulation helps stabilize optimization at high Reynolds numbers by guiding the model toward physically meaningful minima while retaining the physics-based structure of the original PINNs framework.

The relative influence of each loss component was primarily balanced through the selection of the corresponding number of training points associated with the boundary, collocation, and supervised data terms in the residual formulation, rather than through adaptive weighting procedures. In particular, the distribution of collocation points, boundary points, and sparse supervised CFD data was adjusted to ensure stable convergence and prevent any individual loss component from dominating the optimization process. The fixed-weight formulation was adopted in the present study to maintain methodological simplicity and improve reproducibility of the training procedure. Although recent studies have demonstrated that adaptive multi-objective and gradient-based loss-balancing strategies can further improve optimization stability and convergence behavior for stiff PINN formulations~\cite{Bischof2025}, such approaches were not explicitly investigated in the present work and may be explored in future studies.
\subsubsection{Sparse Supervision Strategy for Convection-Dominated Regimes}

The previous low-Reynolds-number investigations demonstrated that physics-only parameterized PINNs can accurately reconstruct laminar incompressible flows when viscous effects remain sufficiently dominant. However, as the Reynolds number increases, the optimization difficulty associated with convection-dominated Navier-Stokes equations increases substantially, leading to degradation in predictive accuracy even when parameterized learning and transfer learning are employed. Consequently, sparse CFD supervision was introduced only for the high-Reynolds-number lid-driven cavity case in order to minimally correct the learned solution manifold while preserving the physics-driven character of the framework.

To assess the practicality and minimum data requirements of the hybrid framework, sparse supervised CFD data were incorporated in two complementary ways. First, although the parameterized hybrid model was trained over the Reynolds-number interval \(500 < Re < 1000\), supervised CFD data were intentionally provided only within the restricted subrange \(750 < Re < 850\). The remaining Reynolds-number space therefore remained effectively unsupervised and was reconstructed primarily through the embedded governing equations and boundary conditions. This localized supervision strategy enables assessment of whether limited high-fidelity information can stabilize PINNs training in convection-dominated regimes without requiring dense labeled datasets across the entire parameter space.

Second, the supervised data fraction \(N_D\) was defined relative to the total number of training points, including collocation and boundary points associated with the medium-resolution computational grid discussed previously. The supervised points were distributed throughout the cavity domain using Monte Carlo sampling in order to provide representative coverage of the primary vortex core, boundary layers, and high-gradient regions near the moving lid and cavity corners.

Table~\ref{tab:sparse_data} summarizes the influence of the supervised data fraction on prediction accuracy for the streamwise velocity component at \(Re=1000\). The results demonstrate that even very limited supervision substantially improves predictive performance compared with purely physics-based high-Reynolds-number training. In particular, increasing the supervised data fraction from \(3\%\) to \(5\%\) reduces the MSE by more than one order of magnitude, from \(8.16\times10^{-3}\) to \(5.15\times10^{-4}\), while simultaneously increasing the \(R^2\) value from \(0.8318\) to \(0.9893\). Beyond approximately \(5\%\) supervision, the improvements become comparatively modest, indicating diminishing returns with increasing data density.

These observations support the central motivation of the present study: at elevated Reynolds numbers, the primary limitation of data-free PINNs is not necessarily the absence of large datasets, but rather insufficient physical guidance in strongly convection-dominated regions of the optimization landscape. Sparse localized supervision therefore acts as a minimal-data correction mechanism that stabilizes optimization and improves reconstruction accuracy without replacing the governing physics as the primary learning constraint. Based on the present results, approximately \(5\%\) sparse CFD supervision was found sufficient to recover accurate high-Reynolds-number flow solutions with strong predictive fidelity and negligible additional computational overhead.

\begin{table}[htbp]
\centering
\caption{Influence of sparse supervised data fraction on prediction accuracy for the streamwise velocity component \(u\) at \(Re = 1000\).}
\label{tab:sparse_data}
\begin{tabular}{c c c}
\hline
\(N_D\) (\% of total training points) & MSE (\(u\)) & \(R^2\) (\(u\)) \\
\hline
3\%  & \(8.16\times10^{-3}\) & 0.8318 \\
\hline
5\%  & \(5.15\times10^{-4}\) & 0.9893 \\
\hline
10\% & \(4.21\times10^{-4}\) & 0.9913 \\
\hline
15\% & \(2.98\times10^{-4}\) & 0.9938 \\
\hline
20\% & \(2.14\times10^{-4}\) & 0.9955 \\
\hline
\end{tabular}
\end{table}
\subsubsection{Contours of \(u\), \(v\), \(p\), and Vorticity}

Contour comparisons of the horizontal velocity (\(u\)), vertical velocity (\(v\)), pressure (\(p\)), and vorticity fields are presented to evaluate the full-field predictive capability of the sparse-supervised hybrid PINNs framework against the high-fidelity \texttt{OpenFOAM~12} reference solutions.

The trained hybrid model was evaluated in both interpolation and extrapolation regimes to examine its generalization capability beyond the supervised Reynolds-number interval. Contour comparisons are therefore presented for \(Re=800\), which lies within the supervised training region, together with \(Re=300\) and \(Re=1200\), which represent extrapolation cases below and above the training range, respectively.

\begin{figure*}[htbp]
    \centering
    \includegraphics[width=\textwidth]{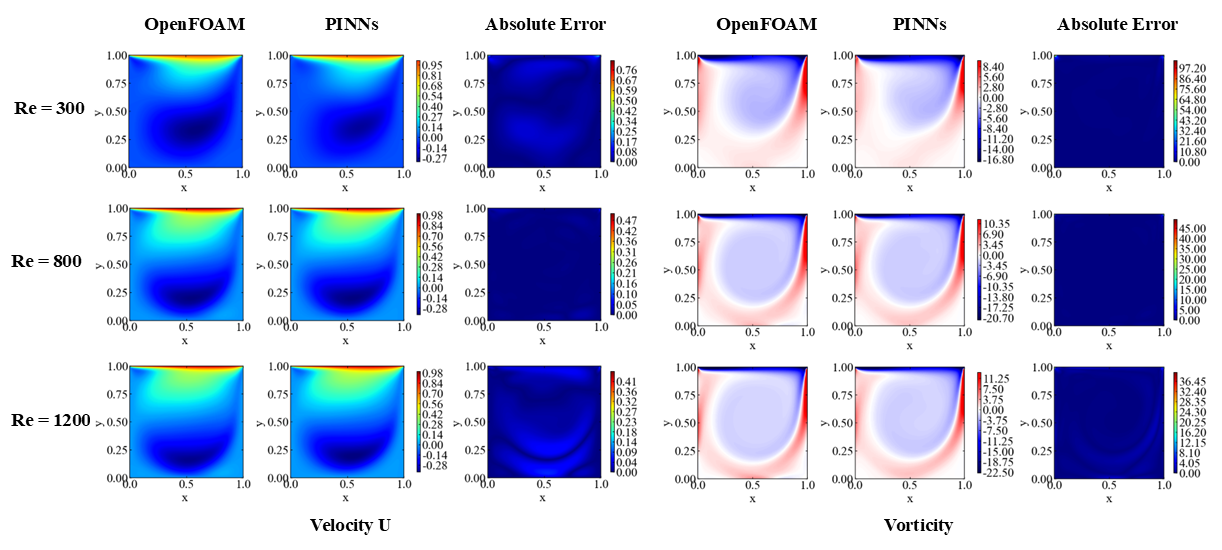}
    \caption{Comparison of horizontal velocity (\(u\)) and vorticity contours between the sparse-supervised hybrid PINNs predictions and OpenFOAM results at \(Re = 300\), \(800\), and \(1200\), together with the corresponding absolute error contours.}
    \label{fig:FigUData1}
\end{figure*}

Figure~\ref{fig:FigUData1} presents comparisons of the horizontal velocity (\(u\)) contours together with the corresponding vorticity contours between the hybrid PINNs predictions and the \texttt{OpenFOAM~12} reference solutions. The associated absolute error distributions are also shown. The hybrid framework accurately reconstructs the dominant recirculating vortex, lid-induced shear layer, and near-wall gradients across all investigated Reynolds numbers. Since vorticity is directly related to velocity gradients, it provides a more stringent assessment of the learned flow physics. The results demonstrate that the sparse-supervised framework successfully captures the primary vortex structures and rotational flow dynamics even in the convection-dominated regime. Although localized discrepancies remain near the moving lid and cavity corners where gradients are strongest, the hybrid PINNs predictions remain in close agreement with the CFD solutions. Importantly, accurate reconstruction is also observed in the extrapolation cases at \(Re=300\) and \(Re=1200\), demonstrating that the hybrid framework retains good predictive capability outside the supervised Reynolds-number interval.

\begin{figure*}[htbp]
    \centering
    \includegraphics[width=\textwidth]{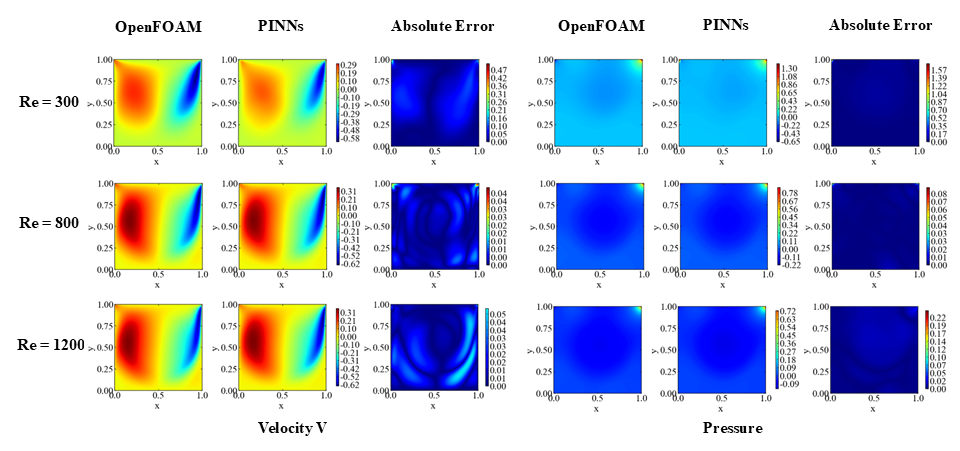}
    \caption{Comparison of vertical velocity (\(v\)) and pressure (\(p\)) contours between the sparse-supervised hybrid PINNs predictions and OpenFOAM results at \(Re = 300\), \(800\), and \(1200\), together with the corresponding absolute error contours.}
    \label{fig:FigVPData1}
\end{figure*}

Similarly, Fig.~\ref{fig:FigVPData1} presents comparisons of the vertical velocity (\(v\)) and pressure (\(p\)) contours between the hybrid PINNs framework and \texttt{OpenFOAM~12}. The pressure field remains accurately reconstructed throughout both interpolation and extrapolation regimes, indicating that the hybrid framework preserves the global pressure dynamics imposed by incompressibility and momentum conservation. The vertical velocity field is also captured accurately overall, although slightly larger localized deviations are observed at \(Re=1200\), reflecting the increased sensitivity of secondary vortical structures and corner eddies in strongly convection-dominated flows.

In summary, the contour comparisons confirm that sparse CFD supervision substantially improves the robustness and predictive capability of PINNs in convection-dominated regimes. In contrast to data-free PINNs, which exhibit noticeable degradation as convection effects intensify, the hybrid framework successfully reconstructs both primary flow variables and derived quantities while maintaining good extrapolation performance at Reynolds numbers outside the supervised training interval.The results further support the interpretation that sparse CFD supervision acts primarily as a corrective stabilization mechanism for high-Reynolds-number optimization rather than as a replacement for physics-based learning.

\subsubsection{Velocity Profiles}

Velocity profiles extracted along selected geometric centerlines of the cavity are used to provide a quantitative assessment of the predictive accuracy of the sparse-supervised hybrid PINNs framework.Velocity profiles are evaluated at Reynolds numbers corresponding to the lower bound, interior region, and upper bound of the high-Reynolds-number training interval, specifically \(Re=500\), \(800\), and \(1000\). This selection enables assessment of the predictive capability of the hybrid framework across the entire trained Reynolds-number range while also evaluating the effectiveness of localized sparse supervision in recovering physically consistent velocity distributions.

\begin{figure}[h]
    \centering
    \includegraphics[width=1.0\linewidth]{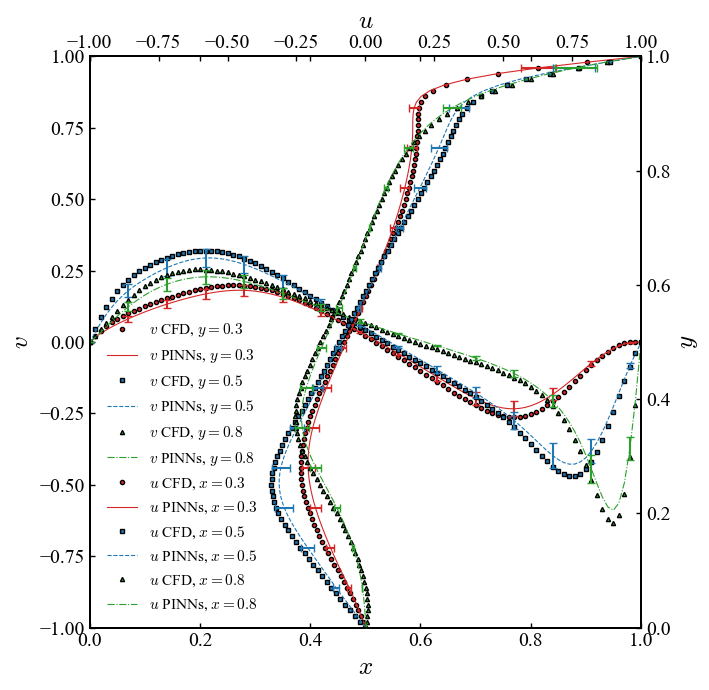}
    \caption{Comparison of \(u\)-velocity profiles along vertical lines (\(x = 0.3\), \(0.5\), \(0.8\)) and \(v\)-velocity profiles along horizontal lines (\(y = 0.3\), \(0.5\), \(0.8\)) between the sparse-supervised hybrid PINNs predictions and OpenFOAM results at \(Re = 500\). Error bars represent \(\pm 10\%\) of the CFD reference values.}
    \label{fig:uvRe500}
\end{figure}

\begin{figure}[h]
    \centering
    \includegraphics[width=1.0\linewidth]{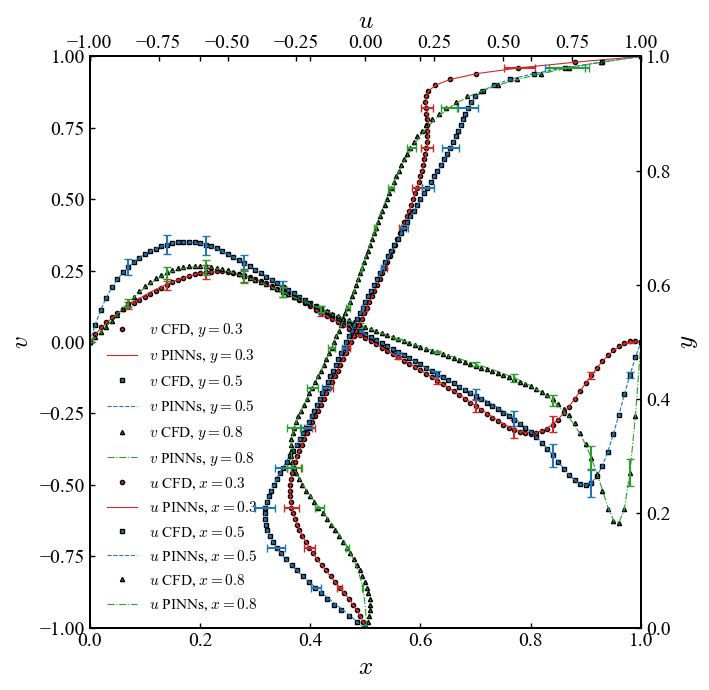}
    \caption{Comparison of \(u\)-velocity profiles along vertical lines (\(x = 0.3\), \(0.5\), \(0.8\)) and \(v\)-velocity profiles along horizontal lines (\(y = 0.3\), \(0.5\), \(0.8\)) between the sparse-supervised hybrid PINNs predictions and OpenFOAM results at \(Re = 800\). Error bars represent \(\pm 10\%\) of the CFD reference values.}
    \label{fig:uvRe800}
\end{figure}

\begin{figure}[h]
    \centering
    \includegraphics[width=1.0\linewidth]{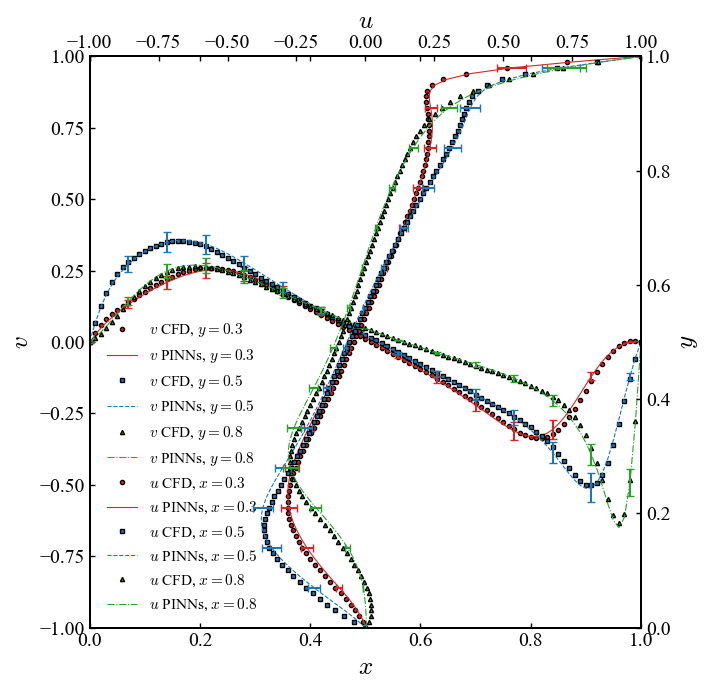}
    \caption{Comparison of \(u\)-velocity profiles along vertical lines (\(x = 0.3\), \(0.5\), \(0.8\)) and \(v\)-velocity profiles along horizontal lines (\(y = 0.3\), \(0.5\), \(0.8\)) between the sparse-supervised hybrid PINNs predictions and OpenFOAM results at \(Re = 1000\). Error bars represent \(\pm 10\%\) of the CFD reference values.}
    \label{fig:uvRe1000}
\end{figure}

Figures~\ref{fig:uvRe500}, \ref{fig:uvRe800}, and \ref{fig:uvRe1000} compare the velocity profiles predicted by the sparse-supervised hybrid PINNs framework against the corresponding \texttt{OpenFOAM~12} reference solutions at \(Re=500\), \(800\), and \(1000\), respectively. The streamwise velocity (\(u\)) profiles are extracted along vertical lines located at \(x=0.3\), \(0.5\), and \(0.8\), while the transverse velocity (\(v\)) profiles are extracted along horizontal lines located at \(y=0.3\), \(0.5\), and \(0.8\).

The error bars shown in the figures correspond to \(\pm 10\%\) of the CFD reference values. Overall, the hybrid PINNs predictions remain within these bounds across most sampled locations and Reynolds numbers, demonstrating strong agreement with the \texttt{OpenFOAM~12} solutions. The agreement is particularly strong within the supervised Reynolds-number interval near \(Re=800\), where the sparse CFD guidance directly assists optimization of the convection-dominated flow structures.

At \(Re=500\), which lies near the lower boundary of the training interval, the velocity profiles remain accurately reconstructed and closely match the reference CFD data. Similarly, accurate agreement is maintained at \(Re=1000\), although slightly larger deviations are observed near regions of elevated shear close to the moving lid and cavity corners, where sharp gradients become increasingly difficult for physics-informed optimization.

The results demonstrate that sparse localized CFD supervision, combined with transfer learning, substantially improves the ability of parameterized PINNs to recover high-Reynolds-number velocity distributions while still preserving the governing equations as the primary learning constraint. These findings further support the interpretation that sparse supervision functions primarily as a minimal-data correction mechanism for convection-dominated PINNs rather than as a fully data-driven replacement for physics-informed learning.

\subsubsection{Integral Force Analysis on the Moving Lid}

To further assess the global predictive capability of the sparse-supervised hybrid PINNs framework, integral force quantities acting on the moving lid were computed and compared against the corresponding \texttt{OpenFOAM~12} reference solutions. The tangential shear force, pressure force, and total resultant force were evaluated using the predicted velocity gradients and mean-corrected pressure fields.

Table~\ref{tab:hybrid_force_comparison} summarizes the comparison of the integral force components for Reynolds numbers within the training range and in extrapolative regimes. Overall, the hybrid PINNs framework demonstrates strong agreement with the CFD solutions across the investigated Reynolds number range. The interpolation case at $\mathrm{Re}=800$, which lies inside the sparse-supervision interval ($750 < \mathrm{Re} < 850$), exhibits particularly high accuracy, with errors below $1\%$ for all force components. Moderate increases in error are observed for extrapolative cases, especially away from the supervised Reynolds-number interval, although the overall force predictions remain physically consistent.

These results demonstrate that localized sparse CFD supervision not only improves local flow reconstruction but also enhances prediction of global momentum-transfer characteristics in convection-dominated flows.

\begin{table*}[htbp]
\centering
\caption{Comparison of integral force components on the moving lid between sparse-supervised hybrid PINNs and OpenFOAM solutions for lid-driven cavity flow.}
\label{tab:hybrid_force_comparison}
\begin{tabular}{c c c c c c c c c c}
\hline
$\mathrm{Re}$ &
$F_{x,\mathrm{CFD}}$ &
$F_{x,\mathrm{PINNs}}$ &
Error (\%) &
$F_{y,\mathrm{CFD}}$ &
$F_{y,\mathrm{PINNs}}$ &
Error (\%) &
$F_{\mathrm{total,CFD}}$ &
$F_{\mathrm{total,PINNs}}$ &
Error (\%) \\
\hline

300  & 0.0686443 & 0.0727761 & 6.02 & 0.0050222 & 0.0053311 & 6.15 & 0.0688277 & 0.0729711 & 6.02 \\

500  & 0.0464974 & 0.0454337 & 2.29 & 0.0050545 & 0.0048591 & 3.87 & 0.0467713 & 0.0456928 & 2.31 \\

800  & 0.0330861 & 0.0331530 & 0.20 & 0.0049748 & 0.0049866 & 0.24 & 0.0334580 & 0.0335259 & 0.20 \\

1000 & 0.0283493 & 0.0280831 & 0.94 & 0.0047885 & 0.0049253 & 2.86 & 0.0287508 & 0.0285118 & 0.83 \\

1200 & 0.0250177 & 0.0240602 & 3.83 & 0.0046447 & 0.0047582 & 2.44 & 0.0254452 & 0.0245262 & 3.61 \\

\hline
\end{tabular}
\end{table*}

\subsubsection{Model Accuracy: Comparison Between CFD and PINNs Predictions}

The predictive accuracy of the sparse-supervised hybrid PINNs framework was quantitatively evaluated by comparing the predicted velocity and pressure fields with high-fidelity CFD solutions obtained using \texttt{OpenFOAM~12}. The assessment was performed over a wide range of Reynolds numbers to evaluate both interpolation performance within the trained parameter space and extrapolation capability beyond the supervised region.

Table~\ref{tab:mse_r2_results} reports the MSE and $R^{2}$ values for the horizontal velocity ($u$), vertical velocity ($v$), and pressure ($p$) fields across Reynolds numbers ranging from $\mathrm{Re}=200$ to $\mathrm{Re}=2000$. The Reynolds number interval $\mathrm{Re}\in[500,1000]$, shaded in gray in the table, corresponds to the Reynolds-number range used during hybrid training. However, sparse CFD supervision was intentionally restricted to the localized interval $750 < \mathrm{Re} < 850$, while the remaining Reynolds numbers within the training range were learned primarily through physics-based constraints and transfer learning.

Within the trained Reynolds-number range, the hybrid PINNs framework demonstrates excellent predictive performance. The MSE values remain consistently low for all three flow variables, while the corresponding $R^{2}$ values remain close to unity, indicating strong agreement with the \texttt{OpenFOAM~12} reference solutions. The highest accuracy is observed near the localized supervised interval around $\mathrm{Re}=800$, where sparse CFD data were incorporated during training. This behavior confirms that even localized supervision can substantially improve optimization stability and solution accuracy in convection-dominated regimes.

Outside the trained Reynolds-number range, the model continues to preserve good predictive capability, although gradual degradation in accuracy is observed as the Reynolds number moves farther away from the training interval. This trend is reflected by increasing MSE values and corresponding reductions in $R^{2}$ scores, particularly at $\mathrm{Re}=1500$ and $\mathrm{Re}=2000$. Nevertheless, the hybrid PINNs framework still retains reasonable extrapolation capability, with the dominant flow structures and overall pressure distributions remaining well captured.

The results further demonstrate the effectiveness of combining transfer learning with localized sparse supervision. Compared with data-free PINNs behavior at high Reynolds numbers discussed previously, the incorporation of limited CFD supervision significantly improves convergence robustness, reduces prediction errors, and stabilizes learning in convection-dominated flow regimes without requiring dense labeled data throughout the entire parameter space.

\begin{figure*}[ht]
    \centering
    \includegraphics[width=0.48\textwidth]{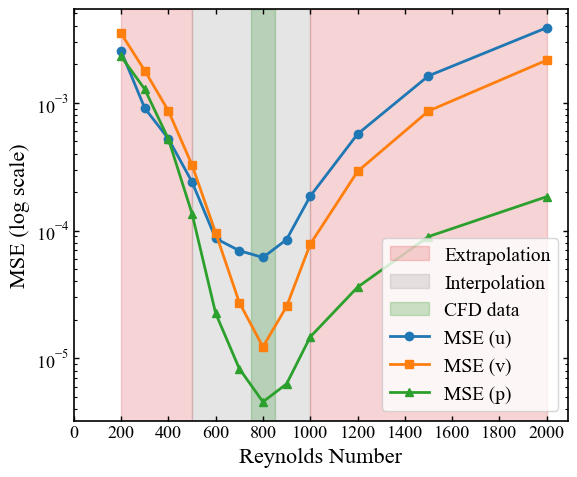}
    \hfill
    \includegraphics[width=0.48\textwidth]{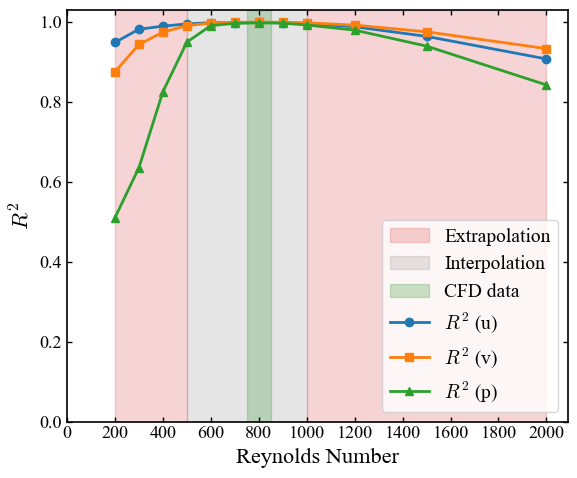}
\caption{
    Quantitative comparison of sparse-supervised hybrid PINNs predictions with CFD reference solutions across Reynolds numbers.
    Left: MSE for the predicted horizontal velocity ($u$), vertical velocity ($v$), and pressure ($p$).
    Right: $R^{2}$ relative to CFD reference solutions.
    The gray-shaded region ($500 \le \mathrm{Re} \le 1000$) denotes the Reynolds-number range used during training, while the green-shaded subregion ($750 < \mathrm{Re} < 850$) indicates the localized Reynolds-number interval where sparse CFD supervision was incorporated. The red-shaded regions represent extrapolation outside the trained Reynolds-number range.
}
\label{fig:mse_r2_combined}
\end{figure*}

Figure~\ref{fig:mse_r2_combined} summarizes these trends by presenting the variation of MSE and $R^{2}$ values with Reynolds number. The gray-shaded region denotes the Reynolds-number range used during training, while the green-shaded subregion indicates the localized interval where sparse CFD supervision was provided. The remaining regions correspond to extrapolative predictions outside the trained parameter space. Overall, the results confirm that the proposed sparse-supervised hybrid PINNs framework provides accurate reduced-order predictions for high-Reynolds-number lid-driven cavity flows while requiring only limited localized CFD assistance.

\begin{table*}[ht]
\centering
\caption{MSE and $R^2$ results for different Reynolds numbers. 
Rows shaded in gray indicate the Reynolds-number range used during hybrid training ($500 \le \mathrm{Re} \le 1000$). Sparse CFD supervision was provided only within the localized interval $750 < \mathrm{Re} < 850$. Unshaded rows correspond to extrapolation outside the trained Reynolds-number range.}
\label{tab:mse_r2_results}
\begin{tabular}{c|ccc|ccc}
\hline
 & \multicolumn{3}{c|}{\textbf{MSE}} & \multicolumn{3}{c}{\boldmath$R^{2}$} \\
\textbf{Re} & $u$ & $v$ & $p$ & $u$ & $v$ & $p$ \\
\hline
200  & $2.54\times10^{-3}$ & $3.49\times10^{-3}$ & $2.31\times10^{-3}$ & 0.9490 & 0.8755 & 0.5103 \\
300  & $9.08\times10^{-4}$ & $1.78\times10^{-3}$ & $1.28\times10^{-3}$ & 0.9818 & 0.9437 & 0.6355 \\
400  & $5.23\times10^{-4}$ & $8.64\times10^{-4}$ & $5.21\times10^{-4}$ & 0.9896 & 0.9746 & 0.8256 \\

\rowcolor{gray!15}
500  & $2.40\times10^{-4}$ & $3.26\times10^{-4}$ & $1.36\times10^{-4}$ & 0.9952 & 0.9909 & 0.9497 \\
\rowcolor{gray!15}
600  & $8.67\times10^{-5}$ & $9.60\times10^{-5}$ & $2.25\times10^{-5}$ & 0.9983 & 0.9974 & 0.9910 \\
\rowcolor{gray!15}
700  & $6.96\times10^{-5}$ & $2.73\times10^{-5}$ & $8.31\times10^{-6}$ & 0.9986 & 0.9993 & 0.9965 \\
\rowcolor{gray!15}
800  & $6.16\times10^{-5}$ & $1.23\times10^{-5}$ & $4.57\times10^{-6}$ & 0.9988 & 0.9997 & 0.9979 \\
\rowcolor{gray!15}
900  & $8.51\times10^{-5}$ & $2.55\times10^{-5}$ & $6.32\times10^{-6}$ & 0.9983 & 0.9993 & 0.9970 \\
\rowcolor{gray!15}
1000 & $1.86\times10^{-4}$ & $7.82\times10^{-5}$ & $1.47\times10^{-5}$ & 0.9962 & 0.9979 & 0.9925 \\

1200 & $5.68\times10^{-4}$ & $2.90\times10^{-4}$ & $3.60\times10^{-5}$ & 0.9880 & 0.9922 & 0.9796 \\
1500 & $1.62\times10^{-3}$ & $8.61\times10^{-4}$ & $8.97\times10^{-5}$ & 0.9640 & 0.9755 & 0.9396 \\
2000 & $3.86\times10^{-3}$ & $2.15\times10^{-3}$ & $1.84\times10^{-4}$ & 0.9075 & 0.9334 & 0.8425 \\
\hline
\end{tabular}
\end{table*}

\subsubsection{Loss Function Convergence, Hyperparameter Tuning, and GPU Utilization}

The incorporation of sparse CFD data in the hybrid PINNs framework introduces a supervised data-loss component in addition to the physics-based residual losses. This hybrid formulation improves convergence stability and prediction accuracy in convection-dominated flow regimes, where purely physics-informed training often suffers from optimization stiffness and imbalance between convective and diffusive residuals.

\begin{figure}[htbp]
    \centering
    \includegraphics[width=0.50\textwidth]{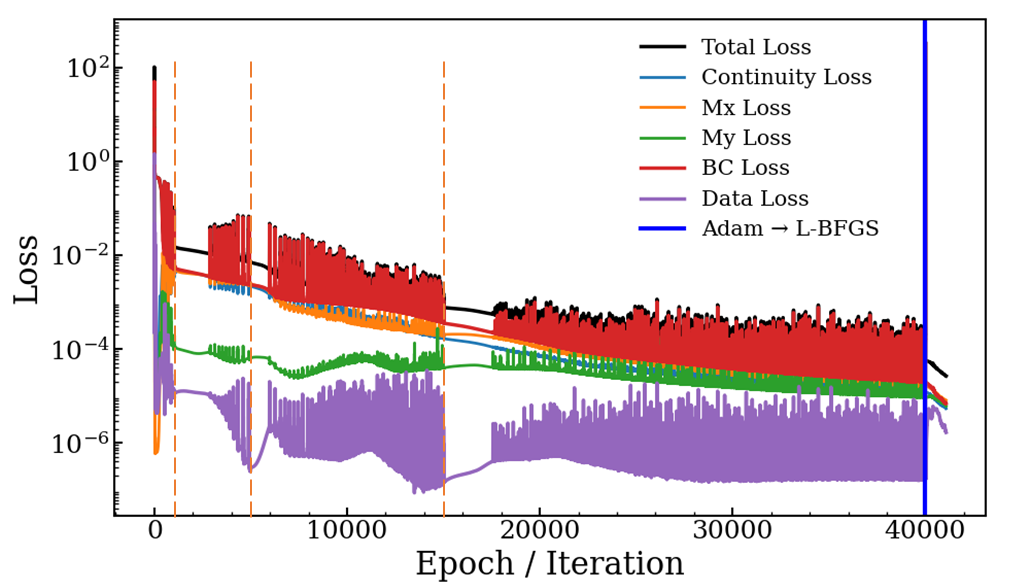}
    \caption{
    Convergence history of the hybrid PINNs training process for the high-Reynolds-number regime using the combined Adam + L-BFGS optimization strategy. The figure shows the evolution of the total loss together with the continuity, momentum, boundary-condition, and supervised data-loss components. The vertical blue line indicates the transition from Adam optimization to L-BFGS fine-tuning, and the dotted orange vertical lines the indiacated the epochs at which the learning rate was adjusted.
    }
    \label{fig:loss_hybrid}
\end{figure}

To improve training robustness for the high-Reynolds-number regime, hyperparameter tuning was performed for the optimizer learning-rate schedule while retaining the same neural-network architecture, \((H,N)=(10,80)\), that was previously employed for the data-free PINNs framework. The initial learning rate was selected as \(5\times10^{-3}\), followed by a piecewise decay strategy with transition epochs at \([1000,\ 5000,\ 15000]\). The corresponding learning-rate values were \([5\times10^{-3},\ 10^{-3},\ 5\times10^{-4},\ 10^{-4}]\). The dotted orange vertical lines in Figure~\ref{fig:loss_hybrid} mark the epochs at which the learning rate was adjusted accordingly. This schedule was found to improve optimization stability by enabling rapid initial convergence while preventing oscillatory behavior during later training stages.
Figure~\ref{fig:loss_hybrid} presents the evolution of the individual loss components during training, including the continuity, momentum, boundary-condition, and supervised data losses. The results show that all residual terms decrease consistently throughout training, while the supervised data loss remains comparatively small, indicating that sparse CFD supervision acts primarily as a corrective stabilization mechanism rather than dominating the optimization process.

\begin{figure}[htbp]
    \centering
    \includegraphics[width=0.50\textwidth]{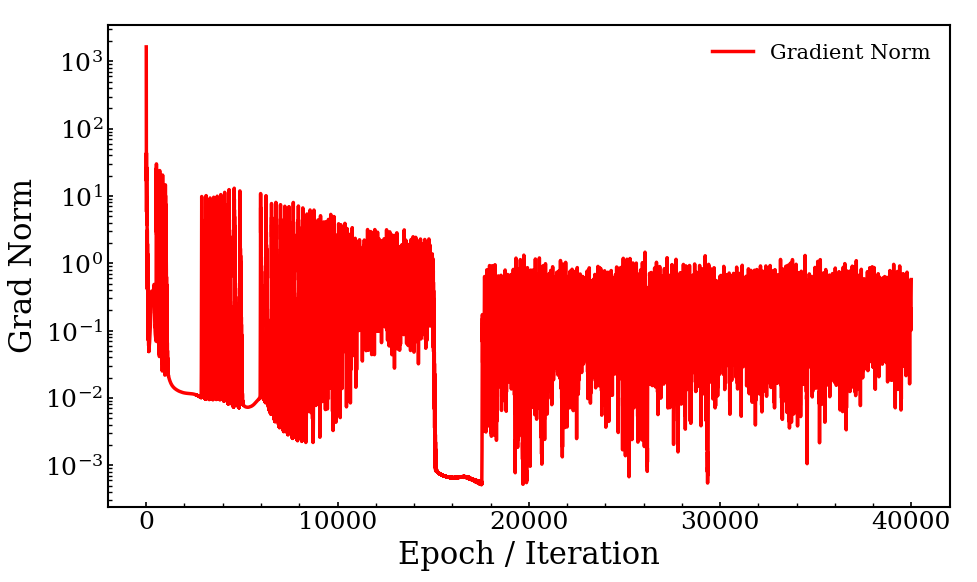}
    \caption{
    Evolution of the gradient norm during hybrid PINNs training for the convection-dominated high-Reynolds-number regime. The gradual stabilization of the gradient norm reflects improved optimization stability after learning-rate decay and late-stage convergence refinement.
    }
    \label{fig:grad_hybrid}
\end{figure}

To further examine optimization behavior, Fig.~\ref{fig:grad_hybrid} shows the gradient norm during training. Large gradient fluctuations are observed during the early stages of optimization due to the strong nonlinearity of convection-dominated flow physics. As training progresses and the learning rate decreases, the gradient norm stabilizes and gradually converges, indicating improved optimization stability and smoother traversal of the loss landscape.

In addition to standard Adam optimization, a hybrid Adam + L-BFGS strategy was investigated. In the present implementation, Adam optimization was first employed for 40,000 epochs to obtain rapid coarse convergence, followed by only 1,000 L-BFGS iterations for fine optimization and local refinement of the solution manifold. The final loss obtained using pure Adam optimization was \(2.06\times10^{-4}\), whereas the combined Adam + L-BFGS strategy reduced the final loss to \(2.67\times10^{-5}\), demonstrating the effectiveness of L-BFGS for late-stage convergence improvement.

Table~\ref{tab:adam_lbfgs_comparison} compares the prediction accuracy obtained using pure Adam optimization and the combined Adam + L-BFGS strategy. The inclusion of L-BFGS consistently improves both the MSE and \(R^2\) metrics across all Reynolds numbers. These results indicate that second-order optimization substantially improves the ability of the PINNs framework to recover high-Reynolds-number flow structures.

Although L-BFGS improves convergence accuracy, it is computationally more expensive and less compatible with efficient GPU parallelization. Unlike Adam, which relies primarily on stochastic first-order gradient updates and benefits strongly from GPU acceleration, L-BFGS requires repeated evaluations of the full-batch loss and approximate Hessian information, resulting in larger memory usage and increased CPU-based computational overhead. In the present study, the 40,000 Adam epochs required approximately 2 GPU hours using an NVIDIA T4 GPU, whereas the additional 1,000 L-BFGS iterations required approximately 1 CPU hour. Despite this additional cost, the improved convergence and accuracy obtained with L-BFGS justify its use as a late-stage fine-tuning optimizer for high-Reynolds-number PINNs.

Additional benchmarking studies were also performed using identical training conditions for both CPU and GPU implementations. In all cases, the same network architecture with \((H,N)=(10,80)\) was employed, and the computational time was measured for every 10,000 training epochs using different grid resolutions. The simulations were performed using an NVIDIA T4 GPU on Google Colab and a 16-core Intel i5 CPU with 64 GB RAM.

Table~\ref{tab:hardware_comparison} summarizes the computational performance comparison between CPU and GPU training. GPU acceleration reduces the training time by approximately a factor of two across all grid resolutions. The computational cost increases with grid density because of the larger number of collocation points and repeated automatic-differentiation operations required for PDE residual evaluation.

Overall, the results demonstrate that proper hyperparameter tuning together with hybrid Adam + L-BFGS optimization substantially improves convergence robustness and prediction accuracy for convection-dominated high-Reynolds-number PINNs, while GPU acceleration remains essential for maintaining practical computational efficiency.

\begin{table}[htbp]
\centering
\caption{Comparison of prediction accuracy between pure Adam optimization and the combined Adam + L-BFGS strategy for the hybrid PINNs framework.}
\label{tab:adam_lbfgs_comparison}
\begin{tabular}{c|cc|cc}
\hline
& \multicolumn{2}{c|}{\textbf{Pure Adam}} & \multicolumn{2}{c}{\textbf{Adam + L-BFGS}} \\
\textbf{Re} & MSE\(_u\) & \(R^2_u\) & MSE\(_u\) & \(R^2_u\) \\
\hline
500  & \(2.42\times10^{-3}\) & 0.9519 & \(2.40\times10^{-4}\) & 0.9952 \\
800  & \(1.49\times10^{-4}\) & 0.9970 & \(6.16\times10^{-5}\) & 0.9988 \\
1000 & \(5.16\times10^{-4}\) & 0.9893 & \(1.86\times10^{-4}\) & 0.9962 \\
1200 & \(2.02\times10^{-3}\) & 0.9572 & \(5.68\times10^{-4}\) & 0.9880 \\
\hline
\end{tabular}
\end{table}

\begin{table}[H]
\centering
\caption{Comparison of PINNs training performance on CPU and GPU hardware using identical network architecture \((H,N)=(10,80)\). Reported times correspond to 10,000 training epochs.}
\label{tab:hardware_comparison}
\begin{tabular}{c c c}
\hline
\textbf{Grid} & \textbf{CPU time (s)} & \textbf{GPU time (s)} \\
\hline
$60 \times 60$ & 2287 & 1173 \\
$80 \times 80$ & 3710 & 1927 \\
$90 \times 90$ & 4460 & 2351 \\
\hline
\end{tabular}
\end{table}
\section{CONCLUSION}

This study systematically investigated a regime-aware sparse-supervised hybrid parameterized physics-informed neural network (PINNs) framework for incompressible Navier-Stokes flows across Reynolds numbers. The Reynolds number was treated explicitly as a continuous network input, enabling a single model to learn a parametric solution manifold spanning multiple flow conditions. Rather than introducing parameterized PINNs themselves, which have been explored in previous studies, the present work focused on understanding the regime-dependent behavior of PINNs and developing a practical physics-first hybridization strategy for convection-dominated incompressible flows.

The results demonstrated that data-free PINNs perform effectively in low-Reynolds-number diffusion-dominated regimes, where smooth solution manifolds and relatively balanced governing equations allow accurate recovery of velocity and pressure fields using only physics-based loss functions. For $\mathrm{Re} \leq 200$, the predicted flow structures, velocity profiles, and pressure distributions showed strong agreement with \texttt{OpenFOAM~12} reference solutions without requiring any CFD supervision.

As the Reynolds number increased and the flow became increasingly convection dominated, the accuracy of data-free PINNs deteriorated substantially due to stronger nonlinear gradients, multiscale flow features, and increasing optimization stiffness. These observations suggest that the effectiveness of PINNs is strongly regime dependent and that physics-only training may become insufficient for accurately recovering high-Reynolds-number flow behavior. In this regime, sparse CFD supervision was introduced not as a primary training mechanism, but as a localized corrective intervention to stabilize optimization and guide the solution toward physically meaningful states.

To address these high-Reynolds-number limitations, a hybrid framework combining transfer learning and localized sparse CFD supervision was developed. Although the overall training range spanned $500 \leq \mathrm{Re} \leq 1000$, CFD supervision was intentionally restricted to the localized interval $750 < \mathrm{Re} < 850$, requiring the model to reconstruct accurate solutions outside the supervised region primarily through embedded physical constraints. The results demonstrated that localized sparse supervision significantly improved convergence and predictive robustness while preserving the physics-informed character of the framework. Quantitative analysis further showed that approximately $5\%$ supervised CFD data were sufficient to recover accurate flow solutions with high predictive fidelity, thereby highlighting the strong data efficiency of the proposed methodology.

The study additionally examined several factors governing the performance and robustness of parameterized PINNs, including sampling strategies, collocation-point density, gradient norm evolution, extrapolation degradation outside the trained Reynolds-number range, and optimization stiffness at elevated Reynolds numbers. These analyses provided additional insight into the practical behavior and limitations of hybrid parameterized PINNs for convection-dominated incompressible flows.

To evaluate generalizability beyond the classical lid-driven cavity benchmark, the framework was further assessed using backward-facing step flow involving separated shear layers and downstream reattachment. The proposed methodology remained robust for this qualitatively different flow topology and accurately captured key flow characteristics, including recirculation zones, shear-layer development, velocity gradients, and reattachment behavior.

Taken together, the present work demonstrates a practical physics-first hybrid PINNs strategy in which pure physics-based learning is employed whenever possible and sparse supervised data are introduced only when necessary to correct high-Reynolds-number failure modes. The results highlight the potential of sparse-supervised hybrid parameterized PINNs for reduced-order modeling and data-assisted simulation of incompressible flows across Reynolds numbers while maintaining substantially lower data requirements than fully supervised approaches. Although the present study focused on steady two-dimensional laminar flows, future work may extend the methodology toward transitional and unsteady flow regimes involving bifurcations, turbulence, multi-parameter systems, and more complex geometries.

\section{REFERENCES}

\nocite{*}
\bibliography{References/aipsamp}
\end{document}